\documentclass[prb, twocolumn, aps,amsmath,amssymb,amsfonts]{revtex4}
\usepackage[german,american,english]{babel}
\usepackage{graphicx}
\usepackage{graphics}
\usepackage{dcolumn}
\usepackage{bm}
\usepackage{amssymb}
\usepackage{amsmath}
\usepackage{amsfonts}
\usepackage{epsfig}

\newcommand {\ket} [1] {| #1 \rangle}

\newcommand {\dbkt} [2] {\langle #1 | #2 \rangle}
\newcommand {\tbkt} [3] {\langle #1 | #2 | #3 \rangle}

 \newcommand {\beq}{\begin{equation}}
\newcommand {\eeq}{\end{equation}}
\begin{document}
\title{Coherent electrical rotations of valley states in Si quantum dots using the phase of the valley-orbit coupling}
\author{Yue Wu and Dimitrie Culcer}
\affiliation{ICQD, Hefei National Laboratory for Physical Sciences at the Microscale, University of Science and Technology of China, Hefei, Anhui 230026, China}
\begin{abstract}
A gate electric field has a small but non-negligible effect on the \textit{phase} of the valley-orbit coupling in Si quantum dots. Finite interdot tunneling \textit{between} valley eigenstates in a double quantum dot is enabled by a small difference in the phase of the valley-orbit coupling between the two dots, and it in turn allows controllable rotations of two-dot valley eigenstates at a level anticrossing. We present a comprehensive analytical discussion of this process, with estimates for realistic structures. 
\end{abstract}
\maketitle

\section{Introduction}
\label{sec:Intro}

Since its inception three decades ago, quantum computing (QC) has developed energetically and spurred incessant innovation, uniting researchers from various areas of physics. \cite{Nielsen_Chuang, LYS_Opt_QC_AP10} One area that has witnessed steady progress in recent years has been quantum control of systems at the nanometer scale, where state of the art experiments have made manipulation of two-level quantum-mechanical systems a reality. At the same time, in addition to accurate and reliable control, a quantum computer requires scalability and long coherence times. This requirement has highlighted solid-state spin systems as a natural choice for quantum bits (qubits).\cite{Kane_Nature98, Loss_PRA98, Burkard_PRB99} Within this area, Si has been known for its extraordinarily long coherence times, \cite{Feher_PR59, Tyryshkin_PRB03, Abe_PRB04, Tyryshkin_JPC06, Hanson_RMP07, Bluhm_LongCoh_10, Tyryshkin_IsoPureSiDnr_T2Secs_11} thanks to the absence of piezoelectric electron-phonon coupling, \cite{Prada_PRB08} weak spin-orbit coupling \cite{Tahan_PRB05, Dery_Si_SO_10}, and nuclear-spin free isotopes, allowing removal of the hyperfine interaction by isotopic purification. \cite{Witzel_AHF_PRB07} As a result, Si spin QC has emerged as an active subfield of modern condensed matter physics. Outstanding experimental progress in Si spin QC has been reported in the last few years in Si quantum dots (QDs), \cite{Lim_SingleElectron_APL09, Nordberg_APL09, Eriksson_QD_SC_APL11, Lim_APL09, MarcusGroup_NatureNano07, Tracy_MOSFET_TunableDQD_APL10, Liu_PRB08, Shaji_NP08, Simmons_DQD_SpinBloc_LET_PRB10, Lai_SiDQD_SpinBlock_10, Borselli_SpinBloc_11, Hayes_09, Xiao_MOS_SpinRelax_PRL10, Thalak_1glShot_10, Morello_1glShot_Nature10, Eriksson_SiQbt_Loading_10, Eriksson_APL10, ChanDzurak_SiQD_Shuttle_APL11, Wild_Si/SiGeQD_10} and in donor-based architectures. \cite{Vrijen_PRA00, Andresen_NanoLett07, Mitic_Nanotech08, Kuljanishvili_NP08, Lansbergen_NP08, Fuhrer_NanoLett09, Stegner_NP06, Calderon_Dnr_Crg_Mismatch_PRB10, Mottonen_PRB10, Morello_PSi_rfDMR_11, Fuechsle_NNano10, Fuechsle_SAT_NNano12, LuBrandt_PSi_FID_11} Theoretical research on Si QDs has also evolved at a brisk pace. \cite{Qiuzi_PRB10, Nielsen_09, Nielsen_Config_10, SDS_SiQD_Hubbard_PRB11, Ramon_2Qbt_11, Raith_SiQD_1e_SpinRelax_PRB11, Wang_SiQD_ST_Relax_PRB10, WangWu_SiDQD_ST_Relax_11, Borhani_2spin_Relax_PRB10, Hu_2spin_e-ph_PRB11, Culcer_APL09, Witzel_EnrichedSi_Decoh_PRL10, Assali_Hyperfine_PRB11} Concomitantly, QDs in other group IV elements are being explored for QC: carbon, \cite{Churchill_PRL09, Palyi_PRB09, Palyi_PRB10, Palyi_PRL11, Wang_GfnQDwSET_APL10, Flensberg_CNT_Bends_PRB10} including nitrogen-vacancy centers in diamond, \cite{ShiDu_DJ_PRL10} and Ge. \cite{Mazzeo_GeQD_10}

Group IV materials (C, Si, Ge) are notable for having equivalent conduction band minima known as valleys. In low-dimensional Si nanostructures typically the two low-lying valleys perpendicular to the interface are important. The valley degree of freedom, investigated many years ago, \cite{Ando_RMP82} has received renewed attention in the context of QC. The multiplicity of the Hilbert space brought about by the existence of equivalent valleys has been shown to hamper \textit{spin} QC. \cite{Koiller_PRL01, Wellard_PRB05, Koiller_PRB06, Culcer_PRB09, Culcer_PRB10} At the same time, the interface potential gives rise to a valley-orbit coupling, which has been studied extensively in recent years, both experimentally \cite{Lai_PRB06, Takashina_PRL06, Goswami_NP07, Lim_SiQD_SpinFill_NT11, Xiao_Valley_APL10, Borselli_Valley_APL11} and theoretically. \cite{Friesen_PRB04, Friesen_PRB07, Friesen_PRB10, Boykin_APL04, Saraiva_PRB09, Saraiva_Extended_PRB10, Nestoklon_PRB06, Srinivasan_APL08, Rahman_Si_EngineerVOS_11} 

Addressing specific valley eigenstates is a profound, challenging and unresolved problem. Despite apparent similarities, valley physics is significantly different from spin-$1/2$ physics, and a pseudospin picture of valley physics is of limited utility. Firstly, no overall conservation law exists for valley composition as for spin. Secondly, a comparison of valley-orbit coupling and spin-orbit coupling sheds additional light on this distinction. In the absence of spin-orbit interactions the one-electron wave-function factorizes into an orbital part and a spin part, whereas no such separation exists for the valley degree of freedom. Unlike the spin-orbit interaction, the valley-orbit coupling cannot be viewed as a wave vector-dependent effective Zeeman field that can tune valley dynamics. Finally, even if the valley splitting is large enough to be resolved, no clear unambiguous signature of individual valleys exists. A \textit{smoking-gun} experiment to identify valley-split states is an ambitious target, and several checks must be made on sets of quantum mechanical states to prove that they are indeed valley-split pairs. For example, for one electron in a single QD it was shown that relaxation between different valley eigenstates proceeds on longer time scales. \cite{Xiao_Valley_APL10} For two electrons in a single QD the ground state moves down as a function of magnetic field,\cite{Lim_SiQD_SpinFill_NT11} and in a double QD (DQD) different valley eigenstates may be identified via resonant tunneling. \cite{Culcer_Roughness_PRB10} 

Since the two valleys are separated by a wave vector of the size of the Brillouin zone, manipulation of the valley degree of freedom is a difficult task. No scheme has been experimentally demonstrated to date for achieving rotations of valley eigenstates in a 2DEG or in a single QD. Nevertheless, a recent work demonstrated that rotations of valley eigenstates can be implemented in a DQD, \cite{Culcer_ValleyQubit_PRL12} in which one can engineer local variations in the \textit{magnitude} of the valley-orbit coupling. This is due to the fact that the magnitude of the valley-orbit coupling in Si can be enhanced by a gate electric field. \cite{Saraiva_PRB09} 

The valley-orbit coupling is a complex number, with a magnitude and a phase. The meaning of the phase is somewhat elusive, and it is not observable in a single QD - one requires at least a DQD to observe phase-related effects. The gate electric field affects both the magnitude and the phase of the valley-orbit coupling, even though its effect on the phase is much smaller. \cite{Saraiva_PRB09} In this work we demonstrate that controllable valley rotations in a DQD can also be accomplished using the \textit{phase} of the valley-orbit coupling. We focus on a different parameter regime from that considered in Ref.~\onlinecite{Culcer_ValleyQubit_PRL12}. One key concept that we exploit is the fact that tunneling between \textit{like} valley eigenstates (i.e. conserving the valley eigenstate index) and between \textit{opposite} valley eigenstates (not conserving the valley-eigenstate index) are not independent processes. Rather, one occurs at the expense of the other, and by controlling the phase of the valley-orbit coupling one can tune the tunneling between these two regimes. Control of the phase of the valley-orbit coupling can be achieved using a gate electric field. We devise an analytical model for the valley-orbit coupling, including the correction due to the gate electric field. We determine expressions for the tunneling parameters between valley eigenstates, and their dependence on the phases of valley-orbit couplings in the two dots, as well as numerical estimates and a feasibility study for experimental implementation. We find that the rotation most likely to succeed experimentally involves the lowest two polarized triplet states in a DQD, which have different valley eigenstate composition. 

The outline of this paper is as follows. In Section \ref{sec:DQD} we review briefly the model of the DQD, presenting the confinement potential and envelope functions, as well as an analytical scheme for calculating the valley-orbit coupling. In Section \ref{sec:Tnl} we focus on the tunneling between valley eigenstates, determining the effect of a gate electric field on the phase of the valley-orbit coupling and its subsequent effect on interdot tunneling. We supply numerical estimates of the change in phase of the valley-orbit coupling and of the intervalley tunneling parameter. Section \ref{sec:Rot} discusses the coherent rotation of valley eigenstates of the two lowest polarized triplets in a DQD, including experimental considerations. We conclude with a summary and outlook in Section \ref{sec:Sum}.

\section{Single and double quantum dots}

\label{sec:DQD}

For consistency, we provide in this section the details of multivalley quantum dots, following previous discussions. \cite{Culcer_PRB10, Culcer_ValleyQubit_PRL12, Culcer_Roughness_PRB10} An electron in a single quantum dot $D$ experiences the potential
\begin{equation}
V_D (x, y, z) = \frac{\hbar^2}{2m^*a^2} \, \bigg[ \frac{(x - x_D)^2 + y^2}{a^2} \bigg] + U_0 \, \theta (z) + eF_Dz.
\end{equation}
The location of the dot is given by $(x_D, 0, 0)$ and its Fock-Darwin radius is $a$, with $m^*$ the Si in-plane effective mass, $F_D$ the interface electric field, and $U_0 \, \theta (z)$ the interface potential, with the interface at $z = 0$ and $\theta (z)$ the Heaviside function. In a multi-valley system, in the effective mass approximation (EMA) the QD wave functions are 
\begin{equation}
D_\xi (x, y, z) = \phi_D(x,y) \, \psi (z) \, u_\xi ({\bm r})\, e^{ik_\xi z}, 
\end{equation}
where $u_\xi ({\bm r})$ is the lattice-periodic Bloch function, the valley index $\xi = \{ z, \bar{z} \}$, and $k_{z, \bar{z}} = \pm k_0$, with $k_0 = 0.85 \, (2\pi/a_{Si})$ and $a_{Si}$ the Si lattice constant. The $z$ and $\bar{z}$ states have a vanishingly small overlap which is neglected. The envelopes $\phi_D (x, y)$ are Fock-Darwin states, $\displaystyle \phi_D (x, y) = [1/(a\sqrt{\pi})] \, e^{-\frac{(x - x_D)^2 + y^2}{2a^2}}$. The solution $\psi(z)$ of the EMA equation for motion in the $z$-direction, perpendicular to the interface,\cite{Saraiva_PRB09, Culcer_Roughness_PRB10} is a variational wave function, \cite{Bastard}
\begin{equation}
\begin{array}{rl}
\displaystyle \psi(z) = M \, e^{\frac{k_b z}{2}} \theta(-z) + N\, (z + z_0) \, e^{\frac{-k_{Si} z}{2}} \theta(z),
\end{array}
\end{equation}
where  $k_b = \sqrt{\frac{2m_bU_0}{\hbar^2}}$, $k_{Si}$ is a variational parameter, and continuity of $\psi(z)$ at $z = 0$ requires $M = Nz_0$. In the basis $\{ D_\xi \}$, the Hamiltonian $H_D$, describing one electron in one multivalley QD, has the form
\begin{equation}\label{H1e1d}
H_D = \varepsilon_D + \begin{pmatrix} 0 & \Delta_D \cr \Delta_D^* & 0 \end{pmatrix},
\end{equation}
with the confinement energy $\varepsilon_D$ and valley-orbit coupling $\Delta_D = |\Delta_D| \, e^{-i\phi_D}$, and $\varepsilon_D \gg |\Delta_D|$. The valley-orbit coupling in dot $D$ is given by \cite{Saraiva_PRB09}
\begin{equation}\label{valley-orbit coupling}
\Delta_D = \tbkt{D_z}{(U_0 \, \theta (z) + eF_Dz)}{D_{\bar{z}}}.
\end{equation}
The eigenstates of $H_D$ are
\begin{equation}
\ket{D_\pm} = \frac{1}{\sqrt{2}} \, (\ket{D_z} \pm e^{i\phi_D} \ket{D_{\bar{z}}}).
\end{equation}
This study will focus on a double quantum dot, with the left dot located at $x_L = -x_0$ and the right dot at $x_R=x_0$. Henceforth we use $D \equiv L, R$ quite generally unless we need to refer specifically to the $L, R$ dots. We assume the presence of a top gate that can be adjusted independently for the $L, R$ dots, thus $F_D$ has different values for the two dots, $F_L$ and $F_R$ respectively. Taking into account these features, the confinement potential for this DQD can be written in the form
\begin{equation}
\arraycolsep 0.3 ex
\begin{array}{rl}
\displaystyle V_{DQD} = & \displaystyle \frac{\hbar^2}{2m^*a^4} \, \bigg\{\mathrm{Min} \, [(x-x_0)^2, (x+x_0)^2] + y^2\bigg\} \\ [3ex]
+ & \displaystyle eEx + eF_Lz + eF_Rz + U_0 \, \theta(z),
\end{array}
\end{equation}
where the in-plane electric field $E$ gives the interdot detuning. The overlap of the single-electron wave functions $l = \dbkt{L_\xi}{R_\xi} \ne 0$, which motivates us to construct orthogonal single-electron wave-functions as in Refs. \ \onlinecite{Burkard_PRB99, Culcer_PRB10}. These wave functions are denoted by $\ket{\tilde{L}_\xi} = \frac{\ket{L_\xi} - g \ket{R_\xi}}{\sqrt{1 - 2lg + g^2}}$ and $\ket{\tilde{R}_\xi} = \frac{\ket{R_\xi} - g \ket{L_\xi}}{\sqrt{1 - 2lg + g^2}}$, where $g = (1 - \sqrt{1 - l^2})/l$, so that $\langle \tilde{L}_\xi | \tilde{R}_\xi \rangle = 0$. Next, we orthogonalize $\ket{L_\pm}$ and $\ket{R_\pm}$ as $\ket{\tilde{L}_\pm} = \frac{\ket{L_\pm} - g \ket{R_\pm}}{\sqrt{1 - 2lg + g^2}}$ and $\ket{\tilde{R}_\pm} = \frac{\ket{R_\pm} - g \ket{L_\pm}}{\sqrt{1 - 2lg + g^2}}$. Henceforth we use the states $\ket{\tilde{D}_\pm}$ and all matrix elements carry a tilde (though matrix elements with and without tildes are numerically almost identical.) In this basis we have a slightly modified valley-orbit coupling $\tilde{\Delta} = \displaystyle \tbkt{\tilde{D}_z}{U_0 \theta(z) + eFz}{\tilde{D}_{\bar{z}}}$ as well as modified confinement energies $\tilde{\varepsilon}_D$. The assumptions underlying this formulation of the problem have been discussed at length in Ref.\ \onlinecite{Culcer_Roughness_PRB10}.

\section{Interdot tunneling between valley eigenstates}
\label{sec:Tnl}

Electrostatic interactions alter the valley-orbit coupling, as was shown in Ref.~\onlinecite{Saraiva_PRB09}. Even though the fractional change in the valley-orbit coupling can be substantial, electrostatic interactions cannot rotate different valley eigenstates into each other on a single QD. Magnetic interactions are much slower than electrostatic ones due to the smallness of the Bohr magneton. Consider a linear magnetic field gradient $\mu_B (\partial B/\partial z) \, z$. The form of the valley-orbit coupling matrix element due to this interaction is the same as the electrostatic potential $eFz$, except it is smaller by $\approx 10^7$ even for the largest magnetic field gradients achievable experimentally. \footnote{We thank Joerg Wunderlich for this information.} Therefore, unsurprisingly, no scheme has been devised for manipulating valley eigenstates via control of the valley-orbit coupling in a single Si QD (but see Ref.~\onlinecite{Palyi_PRL11} for C).

This work is primarily devoted to a new method of harnessing the (small) electric field effect on the phase of the valley-orbit coupling in order to achieve controllable rotations of valley eigenstates in a \textit{double} quantum dot. When the top gate electric field is the same for the $L$ and $R$ dots, so that $F_L = F_R$, the effective confinement potential experienced by the two dots is the same. If, in addition, the interface is sharp along the growth direction and flat perpendicular to it, or if interface roughness is correlated over distances much shorter than the size of the QD, the valley eigenstates $\ket{D_\pm}$ are identical in both dots. \cite{Culcer_Roughness_PRB10} Under such circumstances interdot tunneling occurs only between the same valley eigenstates ($+$ to $+$ and $-$ to $-$), while interdot tunneling between valley eigenstates ($+$ to $-$ and $-$ to $+$) is suppressed. In this section we discuss the modification of the phase of the valley-orbit coupling due to a top-gate, and demonstrate that tuning the top-gate electric field to be different on the $L$ and $R$ dots enables a small amount of interdot tunneling between different valley eigenstates which can be effective in the neighborhood of a level anticrossing.

\subsection{Gate effect on valley-orbit coupling}
\label{sec:Gate}

To determine the valley-orbit coupling $\tilde{\Delta}_D$ of Eq.\ (\ref{valley-orbit coupling}), we expand the lattice-periodic functions $u_\xi ({\bm r})$ as
\begin{equation}
u_\xi ({\bm r}) = \sum_{\bm K} c^\xi_{\bm K} e^{i{\bm K}\cdot{\bm r}},
\end{equation}
with ${\bm K}$ reciprocal lattice vectors. For a perfectly smooth and perfectly sharp interface, neglecting the effect of $F_D$ for the time being, $\tilde{\Delta}_D$  will be referred to as the \textit{global} valley-orbit coupling $\tilde{\Delta}_0$, which can be expressed as
\begin{equation}\label{Delta}
\begin{array}{rl}
\displaystyle \tilde{\Delta}_0 = & \displaystyle U_0 \, N^2 \, z_0^2 \sum_{{\bm K}, Q_z} \, \frac{c^{\xi*}_{{\bm K}} c^{-\xi}_{{\bm K} + Q_z{\bm z}}}{k_b + iq_z},
\end{array}
\end{equation}
where $q_z = Q_z - 2k_0$. For practical purposes, since the Umklapp terms do not contribute, $\Delta_0$ can be accurately approximated by the term with $Q_z = 0$ only
\begin{equation}
\begin{array}{rl}
\displaystyle \tilde{\Delta}_0 \approx & \displaystyle \frac{U_0 \, N^2 \, z_0^2}{k_b - 2ik_0} \, \bigg(\sum_{\bm K} \, c^{z*}_{\bm K} c^{-z}_{\bm K}\bigg) \\
\end{array}
\end{equation}
We note that $z_0$ depends on $k_{Si}$, which in turn depends on the gate electric field. Nevertheless, the effect of the gate on $z_0$ is weak, and $z_0$ is mostly determined by $k_b$, which is fixed by the interface potential.

Our first task is to determine the effect of the top gate $F_D$ on the valley-orbit coupling. We evaluate the contribution due to $F_D$, which arises from the matrix element $eF\tbkt{\tilde{D}_z}{z}{\tilde{D}_{-z}}$,
\begin{widetext}
\begin{equation}
\begin{array}{rl}
\displaystyle \tilde{\Delta}_E = & \displaystyle \tbkt{\tilde{D}_z}{eFz}{\tilde{D}_{-z}} \\ [3ex]
\approx & \displaystyle eF \int_{-\infty}^{\infty} \int_{-\infty}^{\infty} dx\, dy\, |\phi(x,y)|^2  \int_{-\infty}^{\infty} dz\, z \, |\psi (z)|^2 \, e^{-2ik_\xi z}\, u^*_z ({\bm r}) \, u_{-z} ({\bm r}) \\ [3ex]
\approx & \displaystyle eF N^2 \sum_{{\bm K}Q_z} c^{*z}_{{\bm K}} c^{-z}_{{\bm K} + Q_z\hat{\bm z}} \bigg\{z_0^2\int_{-\infty}^0 dz\, z \, e^{(k_b + iq_z) z} + \int_0^{\infty} dz\, z \, (z + z_0)^2 e^{- (k_{Si} - iq_z) z}\bigg\}.
\end{array}
\end{equation}
The integrals are trivial (see Appendix \ref{sec:Int}), and terms with $Q_z \ne 0$ are negligible, yielding
\begin{equation}
\begin{array}{rl}
\displaystyle \tilde{\Delta}_E \approx eF N^2 \sum_{{\bm K}} c^{*z}_{{\bm K}} c^{-z}_{\bm K} \bigg[ - \frac{z_0^2}{(k_b + iq_z)^2} +  \frac{z_0^2}{(k_{Si} - iq_z)^2}  + \frac{4z_0}{(k_{Si} - iq_z)^3} + \frac{6}{(k_{Si} - iq_z)^4} \bigg].
\end{array}
\end{equation}
The full valley-orbit coupling, including the gate correction, can be written as
\begin{equation}
\begin{array}{rl}
\displaystyle \tilde{\Delta}_D \approx & \displaystyle N^2 \bigg(\sum_{\bm K} \, c^{z*}_{\bm K} c^{-z}_{\bm K}\bigg) \bigg\{\frac{U_0z_0^2}{k_b-2ik_0} + eF\bigg[ - \frac{z_0^2}{(k_b + iq_z)^2} +  \frac{z_0^2}{(k_{Si} - iq_z)^2}+ \frac{4z_0}{(k_{Si} - iq_z)^3} + \frac{6}{(k_{Si} - iq_z)^4} \bigg] \bigg\} .
\end{array}
\end{equation}
\end{widetext}
The phase $\tilde{\phi}_D = {\rm arg} \, \tilde{\Delta}_D$ of the valley-orbit coupling
\begin{equation}
\tilde{\phi}_D= \arctan \bigg(\frac{{\rm Im} \, \tilde{\Delta}_D}{{\rm Re} \, \tilde{\Delta}_D}\bigg).
\end{equation}
Assuming that the only difference in $\tilde{\Delta}_D$ between the dots comes from the gate electric field over one dot (which we take to be $L$), so that $\tilde{\Delta}_L = \tilde{\Delta}_0 + \tilde{\Delta}_E$ while $\tilde{\Delta}_R = \tilde{\Delta}_0$, one can write approximately
\begin{equation}
\begin{array}{rl}
\displaystyle \tilde{\phi}_L - \tilde{\phi}_R = & \displaystyle \arctan\bigg(\frac{{\rm Im} \, \tilde{\Delta}_L}{{\rm Re} \, \tilde{\Delta}_L}\bigg) - \arctan\bigg(\frac{{\rm Im} \, \tilde{\Delta}_R}{{\rm Re} \, \tilde{\Delta}_R}\bigg) \\[3ex]
\displaystyle = & \displaystyle \arctan\bigg[\frac{{\rm Im} \, (\tilde{\Delta}_0 + \tilde{\Delta}_E)}{{\rm Re} \, (\tilde{\Delta}_0 + \tilde{\Delta}_E)} \bigg] - \arctan\bigg(\frac{{\rm Im} \, \tilde{\Delta}_0}{{\rm Re} \, \tilde{\Delta}_0} \bigg).
\end{array}
\end{equation}
Experimentally, both top and back gates are required to control the valley-orbit coupling and QD energy levels independently, as discussed extensively in Ref.~\onlinecite{Culcer_ValleyQubit_PRL12}.

\subsection{Tunneling between like and opposite valley eigenstates in a DQD}

Two tunneling parameters are relevant to a multivalley DQD. Tunneling between like valley eigenstates is given by $\tilde{t}_{--}$, which we expect to be dominant given the outstanding quality of present-day Si interfaces.\cite{Lim_APL09, Lai_SiDQD_SpinBlock_10} Tunneling between opposite valley eigenstates is given by $\tilde{t}_{-+}$. This latter parameter is zero if the valley-orbit coupling is exactly the same on both dots. The general formulas for $\tilde{t}_{--}$ and $\tilde{t}_{-+}$ are
\begin{equation}
\begin{array}{rl}
\displaystyle \tilde{t}_{--} = & \displaystyle \frac{\tilde{t}}{2} \, [1 + e^{-i(\tilde{\phi}_L - \tilde{\phi}_R)}] \\ [1ex]
\displaystyle \tilde{t}_{-+} = & \displaystyle \frac{\tilde{t}}{2} \, [1 - e^{-i(\tilde{\phi}_L - \tilde{\phi}_R)}],
\end{array}
\end{equation}
where $\tilde{t} = \tilde{t}_0 + \tilde{s}$, with $\tilde{t}_0 = \tbkt{\tilde{L}_\xi}{H}{\tilde{R}_\xi}$ and the two-particle term $\tilde{s} = \tbkt{L_\xi^{(1)} L_\xi^{(2)}}{V_{ee}}{L_\xi^{(1)}  R_\xi^{(2)}}$, with the superscript $(i)$ denoting the $i$-th electron. The tunneling matrix element $\tilde{t}$, including the Coulomb contribution is calculated in the Appendix. This tunneling parameter is the same for both values of $\xi$. The gate electric field $F_D$ affects both the amplitude and phase of $\tilde{\Delta}_D$, and by tuning $\tilde{\Delta}_D$ to be different on the two dots the tunneling parameters $\tilde{t}_{--}$ and $\tilde{t}_{-+}$ can also be tuned using a gate. 

One can keep $F_R$ constant and tune $F_L$ so that $\tilde{\phi}_L$ is different from $\tilde{\phi}_R$. Assuming this difference to be small, we expand the exponential $e^{-i(\tilde{\phi}_L - \tilde{\phi}_R)}$. The parameter $\tilde{t}_{--}$ acquires a correction linear in the electric field, but it remains effectively $\tilde{t}$, and we shall assume $\tilde{t}_{--} \approx \tilde{t}$ henceforth. However, $\tilde{t}_{-+}$ can be expressed as
\begin{equation}
\begin{array}{rl}
\displaystyle \tilde{t}_{-+} \approx & \displaystyle \frac{i\tilde{t}}{2} \, (\tilde{\phi}_L-\tilde{\phi}_R) \\ [3ex]
\displaystyle \approx & \displaystyle \frac{i\tilde{t}}{2} \, \bigg\{\arctan\bigg[\frac{{\rm Im} \, (\tilde{\Delta}_0 + \tilde{\Delta}_E)}{{\rm Re} \, (\tilde{\Delta}_0 + \tilde{\Delta}_E)} \bigg] - \arctan\bigg(\frac{{\rm Im} \, \tilde{\Delta}_0}{{\rm Re} \, \tilde{\Delta}_0} \bigg)\bigg\}.
\end{array}
\end{equation}
By varying the gate electric field on one of the two QDs, $\tilde{t}_{-+}$ can be tuned from zero to the desired time scale. Experimentally, $\tilde{t}$ and $\tilde{t}_{-+}$ can be identified using resonant tunneling, \cite{Escott_ResTnl_Nanotech10} as described in detail in Ref. \onlinecite{Culcer_Roughness_PRB10}.

\begin{figure}[tbp]
\includegraphics[width=\columnwidth]{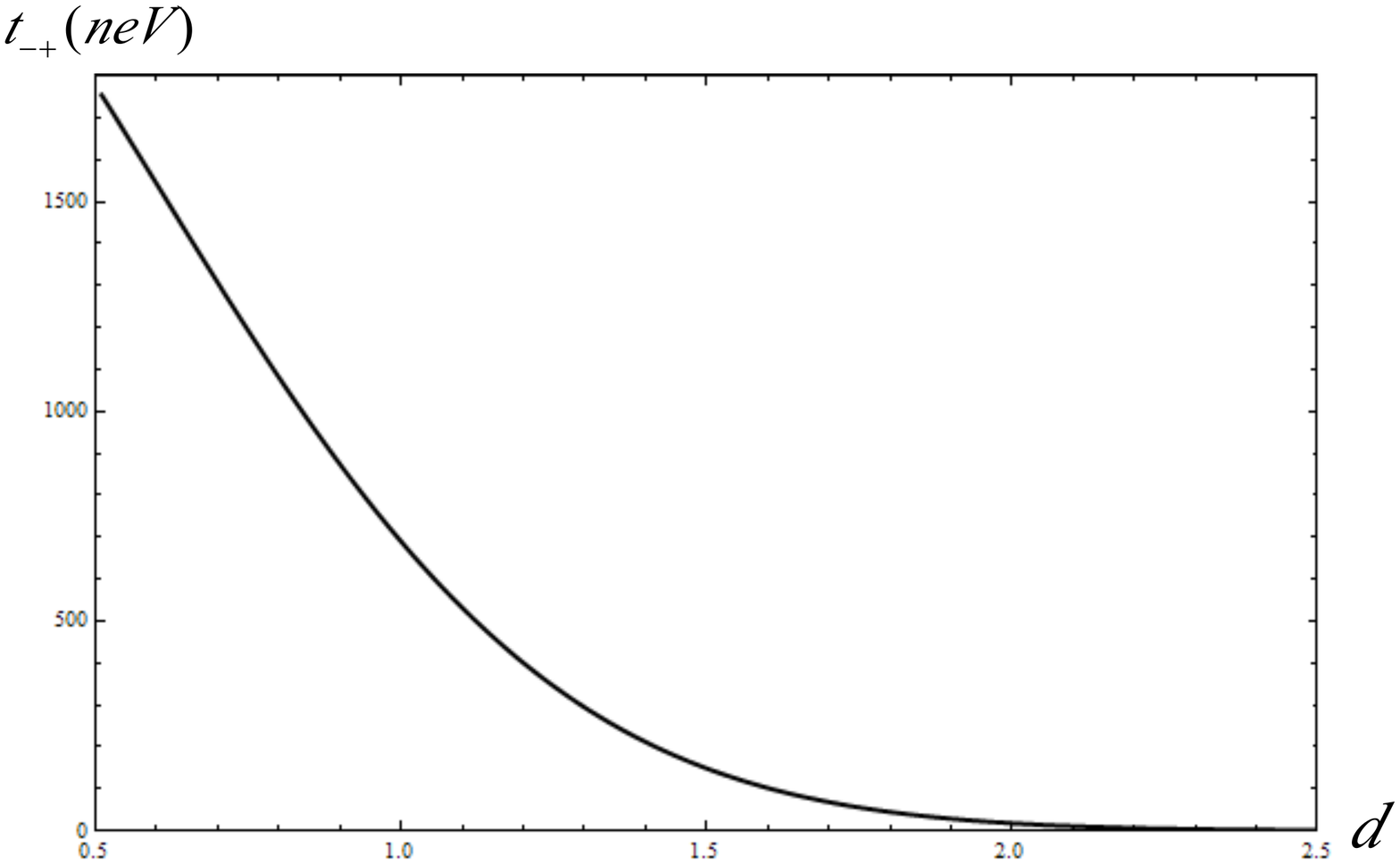}
\caption{Interdot tunneling between valley eigenstates for Si/SiO$_2$. The parameters used are $m_b=0.4m_0, U_0=3.0eV, a=8.2nm$ and $b=1.060nm$.}
\label{t-+}
\end{figure}

We discuss next a series of numerical estimates for the valley-orbit coupling, the effect of an electric field on it, and the interdot tunneling parameter $\tilde t_{-+}$. The dimensionless quantity $d = X_0/a$ represents half the interdot separation in units of the QD radius. A realistic interfacial electric field of 150 kV/cm is assumed to obtain numerical estimates. For a Si/SiO$_2$ interface, with $U_0 \approx$ 3eV and $m_b=0.4m_0$, $b = 1/k_{Si}$ is optimized at 1.06 nm, and we find $|\tilde{\Delta} _0| \approx 230\mu$eV. For a Si/SiGe interface, with $U_0 \approx 150meV$ and $m_b = m_0$, $b$ is optimized at $1.217 nm$, and we find $|\tilde{\Delta} _0| \approx 200 \mu eV$. Table \ref{table:Delta} displays $\tilde{\Delta}_0$, $\tilde{\Delta}_E$ and $\tilde t_{-+}$ for experimentally realistic parameters, and the parameter $\tilde t_{-+}$ is plotted in Figure \ref{t-+}. The key effect investigated in this work concerns the phase change due to $\tilde{\Delta}_E$, which enables tunneling between opposite valley eigenstates. Determination of $\tilde t_{-+}$ for experimentally relevant parameters yields a time scale of $0.8-3.5\mu s$, which can be easily accessed in the laboratory. This demonstrates the feasibility of two-dot manipulation.

\begin{table}
\label{table:Delta}
\caption{\label{table:Delta} Numerical estimates of $\Delta_0$, $\Delta_E$ and $\tilde t_{-+}$ in Si QDs.}
\begin{tabular}{|c|c|c|c|}\hline
   &   & \multicolumn{2}{c|}{Si/SiO$_2$} \\\cline{3-4}
\raisebox{1.5ex}[0pt]{Material} & \raisebox{1.5ex}[0pt]{Si/SiGe} & {m$_b$=0.4m$_0$} & {m$_b$=0.3m$_0$} \\\hline
$|\Delta_0|(\mu eV)$ & 200 & 230 & 150 \\\hline
$|\Delta_E|(\mu eV)$ & $1.7\times10^{-2}$ & $6.9\times10^{-2}$ & $7.9\times10^{-2}$ \\\hline
$|\tilde{\phi}_L-\tilde{\phi}_R|(rad)$ & $8.6\times10^{-5}$ & $2.9\times10^{-5}$ & $1.2\times10^{-4}$ \\\hline
$|\tilde{t}_{-+}| (neV)(d=2.5)$ & 0.28 & 0.20 & 0.84 \\\hline
$|\tilde{t}_{-+}| (\mu s)$ & 2.3 & 3.3 & 0.8 \\\hline
\end{tabular}
\end{table}

We note in closing this section that the phase of the valley-orbit coupling cannot be of use in the case of a single QD, regardless of its occupation number. In fact the phase of the valley-orbit coupling cannot be measured in a single QD. Both the confinement energy $\tilde \varepsilon_0 = \tbkt{\tilde{D}_z}{T+V_D}{\tilde{D}_{\bar{z}}}$ and the on-site Coulomb energy $\tilde u$ are independent of the phase of $\tilde{\Delta}$. To extract information about the phase of the valley-orbit coupling the minimal requirement is a DQD, in which the phase difference between the two dots gives rise to the measurable quantity $\tilde t_{-+}$, as outlined above. 

\section{Coherent rotations of two-electron states in a double quantum dot}
\label{sec:Rot}

Two-electron states may be spin singlets or spin triplets. Previous work has found that the two lowest energy spin-singlet levels do not cross as a function of detuning. \cite{Culcer_Roughness_PRB10} Even for the case in which the lowest singlet state is easily initialized, no value of the detuning exists where one can controllably induce mixing between this state and another valley eigenstate of different valley composition. Therefore we focus on the spin triplet branch, in which it is easiest to access the polarized spin-down triplet states by applying a magnetic field of order 1-2 T. When the lowest-energy spin-polarized triplet state is initialized, the detuning can be swept to an anticrossing in the triplet energy-level spectrum, where one can mix two spin-polarized triplet states with different valley composition. This provides a scheme for the implementation of coherent rotations of valley eigenstates, as we will see below.

\begin{figure}[tbp]
\includegraphics[width=\columnwidth]{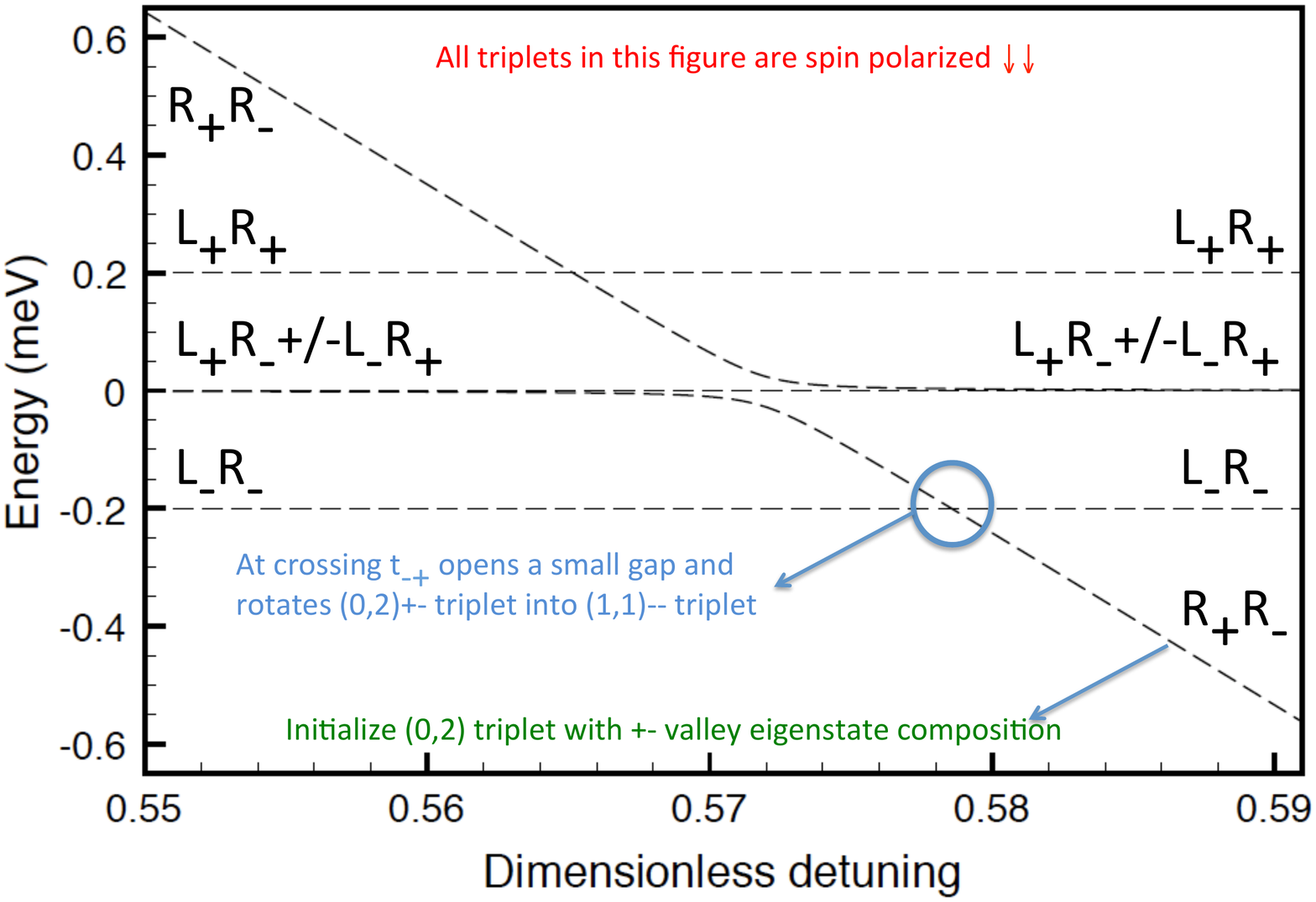}
\caption{Triplet levels in a double quantum dot in the absence of interface roughness, plotted as a function of the detuning, which has been suitably non-dimensionalized as in Ref.\ \onlinecite{Culcer_PRB10}. We have assumed a bare valley splitting $\tilde{\Delta}_0$ of 0.1meV.}
\label{Triplets}
\end{figure}

With the magnetic field set to zero for the time being, we investigate the dynamics of the lowest two triplet states. The difference in $|\tilde{\Delta}_D|$, i.e. $|\tilde{\Delta}_L| - |\tilde{\Delta}_R|$, induced by the gate electric field can be ignored for our purposes. We therefore write $|\tilde{\Delta}_L| = |\tilde{\Delta}_R| = |\tilde{\Delta}|$. We use the basis $\{\tilde{T}^{LR}_{- -}, \tilde{T}^{LR}_{++}, \tilde{T}^{LR}_{+-}, \tilde{T}^{LR}_{-+}, \tilde{T}^{RR}_{+-}\}$ of Ref. \onlinecite{Culcer_PRB10}
\begin{equation}
\begin{array}{rl}
\displaystyle \tilde{T}^{LR}_{- -}= & \displaystyle \frac{1}{\sqrt 2}(\tilde L^{(1)}_- \tilde R^{(2)}_- - \tilde L^{(2)}_- \tilde R^{(1)}_-) \\[3ex]
\displaystyle \tilde{T}^{LR}_{++} = & \displaystyle \frac{1}{\sqrt 2}(\tilde L^{(1)}_+ \tilde R^{(2)}_+ - \tilde L^{(2)}_+ \tilde R^{(1)}_+) \\[3ex]
\displaystyle \tilde{T}^{LR}_{+-} = & \displaystyle \frac{1}{\sqrt 2}(\tilde L^{(1)}_+ \tilde R^{(2)}_- - \tilde L^{(2)}_+ \tilde R^{(1)}_-) \\[3ex]
\displaystyle \tilde{T}^{LR}_{-+} = & \displaystyle \frac{1}{\sqrt 2}(\tilde L^{(1)}_- \tilde R^{(2)}_+ - \tilde L^{(2)}_- \tilde R^{(1)}_+) \\[3ex]
\displaystyle \tilde{T}^{RR}_{+-} = & \displaystyle \frac{1}{\sqrt 2}(\tilde R^{(1)}_+ \tilde R^{(2)}_- - \tilde R^{(2)}_+ \tilde R^{(1)}_-).
\end{array}
\end{equation}
Barring a constant offset and noting that $\tilde{t}_{+-} = \tilde{t}_{-+}$, in this basis the Hamiltonian is
\begin{equation}
\arraycolsep 0.3 ex
\begin{array}{rl}
\displaystyle \tilde{H}_T = & \displaystyle \begin{pmatrix}
- 2|\tilde{\Delta}| & 0 & 0 & 0 & -\tilde{t}_{-+} \cr
0 & 2|\tilde{\Delta}| & 0 & 0 & \tilde{t}_{-+} \cr
0 & 0 & 0 & 0 & \tilde{t} \cr
0 & 0 &  0 & 0 & - \tilde{t} \cr
-\tilde{t}_{-+}^* & \tilde{t}_{-+}^* & \tilde{t} & - \tilde{t} & - \tilde{\delta}
\end{pmatrix}
\end{array}
\end{equation}
The detuning is defined as is defined as $\tilde{\delta} = (\tilde{\varepsilon}_L - \tilde{\varepsilon}_R) -  (\tilde{u} -
\tilde{k})$, where $\tilde{u}$ is the on-site Coulomb repulsion and $\tilde{k}$ the two-electron two-dot direct Coulomb integral, both discussed in Ref.\ \onlinecite{Culcer_PRB10}. We diagonalize this Hamiltonian treating $\tilde{t}_{+-}$ as a perturbation. The eigenstates without $\tilde{t}_{+-}$ are $\tilde{T}^{LR}_{\pm\pm}$ and
\begin{equation}
\begin{array}{rl}
\displaystyle \tilde{T}^>_{+-} = & \displaystyle \frac{\tilde{\varepsilon}^>_0}{\sqrt{\tilde{\varepsilon}^{>2}_0 + 2\tilde{t}^2}} \, \bigg( \frac{\tilde{t}}{\tilde{\varepsilon}^>_0} \, \tilde{T}^{LR}_{+-} - \frac{\tilde{t}}{\tilde{\varepsilon}^>_0} \, \tilde{T}^{LR}_{-+} + \tilde{T}^{RR}_{+-} \bigg) \\ [3ex]
\displaystyle \tilde{T}^<_{+-} = & \displaystyle \frac{\tilde{\varepsilon}^<_0}{\sqrt{\tilde{\varepsilon}^{<2}_0 + 2\tilde{t}^2}} \, \bigg( \frac{\tilde{t}}{\tilde{\varepsilon}^<_0} \, \tilde{T}^{LR}_{+-} - \frac{\tilde{t}}{\tilde{\varepsilon}^<_0} \, \tilde{T}^{LR}_{-+} + \tilde{T}^{RR}_{+-} \bigg) \\ [3ex]
\displaystyle \tilde{T}^{sym}_{+-} = & \displaystyle \frac{1}{\sqrt{2}} \, \bigg( \tilde{T}^{LR}_{+-} + \tilde{T}^{LR}_{-+} \bigg).
\end{array}
\end{equation}
The notation used above is
\begin{equation}
\begin{array}{rl}
\displaystyle \tilde{\varepsilon}^{>}_0 = \frac{ -\tilde{\delta} + \sqrt{\tilde{\delta}^2 + 8\tilde{t}^2} }{2} \\ [3ex]
\displaystyle \tilde{\varepsilon}^{<}_0 = \frac{ -\tilde{\delta} - \sqrt{\tilde{\delta}^2 + 8\tilde{t}^2} }{2}.
\end{array}
\end{equation}

Reliable preparation of the state $\tilde{T}^<_{+-}$ is accomplished by raising the detuning $\tilde{\delta}$ so that the right dot is much lower in energy than the left. A magnetic field of 1-2 T is applied to separate the spin-polarized triplets from the other two-electron states. At this point two electrons can be unambiguously initialized on the right dot. This has already been demonstrated experimentally. \cite{Lim_SiQD_SpinFill_NT11} (When a magnetic field is applied, the orbital states do change, but the effect will be the same for all triplet states -- there will be a constant offset, which does not affect the argument presented here.)

The energies of the spin triplet states as a function of detuning are plotted in Fig. \ref{Triplets}. The lowest energy states are $\tilde{T}^<_{+-}$ and $\tilde T^{LR}_{--}$. Referring to Fig.~\ref{Triplets}, one now sweeps the detuning until the two states $\tilde{T}^<_{+-}$ and $\tilde{T}^{LR}_{--}$ cross. At the point when they cross, $\tilde{T}^<_{+-} \approx \tilde{T}^{RR}_{+-}$. Thus the matrix element mixing the lowest two triplet states is effectively $\tilde{t}_{+-}$, which gives an anticrossing at this point. Since usually $\tilde{\Delta} \gg \tilde{t}$, the detuning at the anticrossing is approximately given by $\tilde{\delta} \approx 2 \tilde{\Delta}$ (corresponding to the energy difference of the two triplet states when $\tilde{\delta} \ll 0$). The tunneling matrix element between $\tilde{T}^<_{+-}$ and $\tilde T^{LR}_{--}$ at the anticrossing is
\begin{equation}
\begin{array}{rl}
\displaystyle \langle \tilde T^{LR}_{--} | \tilde H_T | \tilde{T}^<_{+-} \rangle
\displaystyle = & \displaystyle \frac{\tilde\varepsilon^<_0}{\sqrt{\tilde\varepsilon^{<2}_0 + 2\tilde{t}^2}} \langle \tilde T^{LR}_{--} | \tilde H_T | \tilde{T}^{RR}_{+-} \rangle \\[3ex]
\displaystyle = & \displaystyle \bigg( \frac {\tilde\delta + \sqrt{\tilde\delta^2+8\tilde t^2}} {\sqrt {2\tilde\delta^2+ 2\tilde\delta \sqrt{\tilde\delta^2+8\tilde t^2} + 16\tilde t^2}}\bigg) \, \tilde t_{-+}. \\[3ex]
\end{array}
\end{equation}
Since $\tilde{\delta} \approx 2\tilde \Delta \gg \tilde t$, the above tunneling matrix element is approximately $\tilde t_{-+}$. The width of the anticrossing is $2\sqrt{2}\tilde t_{-+}$. At the anticrossing point the two states, $\tilde{T}^<_{+-}$ and $\tilde T^{LR}_{--}$, can be rotated into each other on a time scale given by $\hbar/\tilde t_{-+}$, which constitutes a coherent rotation of the valley state of two electrons. The values of $\tilde t_{-+}$ given in Table~\ref{table:Delta}, which are of the order of $\mu$s, can be regarded as maximum values for the given parameters. Tuning the gate electric fields $F_L$ and $F_R$ allows one to tune $\tilde t_{-+}$ from zero to the maximum achievable value, controlling the time scale of the rotation. 

Experimentally one needs to know the valley-orbit coupling $\tilde{\Delta}$ in order to determine where the anticrossing occurs. For this scheme to work it is necessary for the valley-orbit coupling to be measured beforehand.\cite{Thalak_1glShot_10, Lim_SiQD_SpinFill_NT11} Interface roughness gives an additional electric field, but that is a static offset. Finally, applying a large gate electric field may modify the number of electrons in the dot, thus this proposal requires a depletion dot, rather than an accumulation dot.

The discussion above has focused on the implementation of a coherent rotation of valley eigenstates, which in the language of quantum computing would constitute a $\sigma_x$ gate. It is evident that the detuning $\tilde{\delta}$ could provide a $\sigma_z$ gate. It would appear that the method proposed here for rotating valley eigenstates has the advantage of longer pulses, which are better experimentally being more precise. Nevertheless, appealing as it is, the scheme presented in this work cannot be promoted to a qubit as it stands, the main concern in quantum computing applications being dephasing. Longer pulses require a longer coherence time $T_2^*$(as found in the context of the hyperfine interaction in GaAs. \cite{Ribeiro_ST+_10}) We have assumed the hyperfine interaction to eliminated through isotopic purification, phonons to be frozen out at dilution refrigerator temperatures, while interface roughness and screened charged impurities give rise to a constant offset in the QD confinement potential. This leaves the biggest source of dephasing as charge noise due to dangling bonds in the vicinity of the interface, which cause fluctuations in the detuning, and give rise to random telegraph and $1/f$ noise. \cite{Culcer_APL09} Near the anticrossing, where the coherent valley rotation is performed, the energy splitting between the two levels is approximately linear in $\tilde{\delta}$, thus the system is susceptible to charge noise [this term $\propto \tilde{\delta}$ arises from matrix elements of the form $\tbkt{\tilde{L}_-}{V_{DQD}}{\tilde{L}_-}$ and $\tbkt{\tilde{R}_\pm}{V_{DQD}}{\tilde{R}_\pm}$, i.e. one of the states has occupation $(1,1)$ whereas the other has occupation $(0,2)$.] The absence of a sweet spot in the qubit energy splitting does not allow one to minimize the sensitivity to noise. Therefore, unless this noise can be filtered out the coherence time $T_2^*$ may be too short for quantum computation. Reliable numbers for $1/f$ noise in single and double Si QDs are scarce, though $1/f$ noise and the charge offset drift has been measured in SETs \cite{Zimmerman_LongTermCOD_JAP08} and are indicative of the results to be expected in QDs. At the same time, there has been progress of late in combating the effect of noise in QD spin qubits, such as a singlet-triplet qubit via dynamical decoupling, \cite{Barthel_InterfaceDD_PRL10} composite pulses, \cite{Wang_CompPuls_12} and by growing a buried quantum dot \cite{Lu_Si/SiGe_Buried_11} further from the interface.

A recent publication \cite{Culcer_ValleyQubit_PRL12} has demonstrated that quantum bits can be implemented using the valley degree of freedom in Si QDs. One common feature of this work and Ref.\ \onlinecite{Culcer_ValleyQubit_PRL12} is their reliance on local control of the top gate over one quantum dot in order to generate a difference in the valley-orbit coupling $\tilde{\Delta}_E$ for the two quantum dots in order to mix two polarized triplets with different valley composition. Nevertheless, the two schemes employ different principles and are operated in different parameter regimes. In Ref.\ \onlinecite{Culcer_ValleyQubit_PRL12} it is the difference in the magnitude $|\tilde{\Delta}_E|$ of the valley-orbit coupling that is exploited in order to mix the two triplet states $\tilde{T}^{sym}_{+-}$ and $\tilde{T}^{anti}_{+-}$ in the far detuned regime $\tilde{\delta} \ll 0$. In contrast, this work exploits the difference in the phase of $\tilde{\Delta}_D$ in order to mix $\tilde{T}^{RR}_{+-}$ and $\tilde{T}^{LR}_{--}$ at a point where they cross, and it is not $\tilde{\Delta}_E$ itself that causes the mixing but the tunneling between valley eigenstates that it gives rise to. The conclusion that emerges from these two works is the following: the electric field changes both the magnitude and the phase of $\tilde{\Delta}$. When one uses the change in magnitude of $\tilde{\Delta}$, the change in phase is irrelevant. When one uses the change in phase of $\tilde{\Delta}$, the change in magnitude is irrelevant.

\section{Summary and Outlook}
\label{sec:Sum}

We have demonstrated that controllable rotations between valley eigenstates can be performed in double quantum dots in Si using gate electric fields. Finite interdot tunneling \textit{between} valley eigenstates is enabled by a small difference in the \textit{phase} of the valley-orbit coupling in the two dots, and it in turn allows controllable rotations of two-dot valley eigenstates states.

The scheme we propose has relied on a DQD. The possibility exists that extensions may be devised for systems of three QDs, which have also been studied recently. \cite{Laird_ExchQbt_PRB10} Such extensions may uncover new and interesting physics and will be addressed in the future. Furthermore, the scheme we propose is expected to have analogs in carbon nanotube and graphene QDs \cite{Churchill_PRL09, Wang_GfnQDwSET_APL10} as well as recently realized Ge QDs. \cite{Mazzeo_GeQD_10} 

This work is supported by LPS-NSA-CMTC and by the National Natural Science Foundation of China under grant number 91021019. We thank N.~M.~Zimmerman for pointing out the role of charge noise in this system. We gratefully acknowledge discussions with S.~Das Sarma, Andre Saraiva, H.~W.~Jiang, Andr\'as P\'alyi, Wang Yao, R.~J.~Joynt, Andrea Morello, Andew Dzurak, G.~P.~Guo, M.~Xiao, Matthew House, Joerg Wunderlich, Xuedong Hu, Belita Koiller, Ted Thorbeck, M.~A.~Eriksson, and J.~M.~Taylor. 

\appendix

\begin{widetext}

\section{Interdot tunneling parameter $\tilde{t}$}
\label{app:Coulomb}

The interdot tunneling parameter $\tilde{t} = \tilde{t}_0 + \tilde{s}$. The single-particle part of the tunneling parameter, $\tilde{t}_0$, does not depend on the form of the $z$-wave function, and is the same as that calculated in our previous papers (Ref.\ \onlinecite{Culcer_PRB10}). We therefore require the Coulomb integrals for the tunneling parameter. We wish to evaluate the integral
\begin{equation}
I_\psi ({\bm P}, {\bm Q}) = \frac{1}{\pi^2a^4}\int d^3r_1\int d^3r_2 \frac{e^{-\frac{(x_1 - P_x)^2 + (y_1 - P_y)^2}{a^2})} e^{-\frac{(x_2 - Q_x)^2 + (y_2 - Q_y)^2}{a^2})} \psi(z_1)^2\psi(z_2)^2}{|{\bm r}_1 - {\bm r}_2|},
\end{equation}
where $\psi(z)$ is real. Define
\begin{equation}
\arraycolsep 0.3 ex
\begin{array}{rl}
\displaystyle f ({\bm r}) = & \displaystyle \int \frac{d^3k}{(2\pi)^3} \, e^{-i{\bm k}\cdot{\bm r}}F({\bm k}) \\ [3ex]
\displaystyle F ({\bm k}) = & \displaystyle \int d^3r \, e^{i{\bm k}\cdot{\bm r}}f({\bm r}).
\end{array}
\end{equation}
The Fourier transform of $f(\bm r) = 1/r$ is $F(\bm k) = 4\pi/k^2$. We write the term $1/|{\bm r}_1 - {\bm r}_2|$ as a Fourier expansion
\begin{equation}
\frac{1}{|{\bm r}_1 - {\bm r}_2|} = \frac{4\pi}{(2\pi)^3}\int d^3 k_2 \, \frac{e^{-i{\bm k}_2\cdot({\bm r}_1 - {\bm r}_2)}}{k_2^2} =
\frac{1}{2\pi^2}\int d^3 k_2 \, \frac{e^{-i{\bm k}_2\cdot({\bm r}_1 - {\bm r}_2)}}{k_2^2}.
\end{equation}
We Fourier transform the $x$ and $y$-dependent terms
\begin{equation}
\arraycolsep 0.3 ex
\begin{array}{rl}
\displaystyle e^{-\frac{({\bm r}_{1\perp} - {\bm P}_\perp)^2}{a^2}} = & \displaystyle \frac{a^2}{4\pi} \int \!\!\!\!\!  \int d^2k_{1\perp}
e^{-i{\bm k}_{1\perp}\cdot({\bm r}_{1\perp} - {\bm P}_\perp)} e^{- \frac{a^2k_{1\perp}^2}{4}} \\ [3ex]
\displaystyle e^{-\frac{({\bm r}_{2\perp}
- {\bm Q}_\perp)^2}{a^2}} = & \displaystyle \frac{a^2}{4\pi} \int \!\!\!\!\! \int d^2k_{3\perp} e^{-i{\bm k}_{3\perp}\cdot({\bm r}_{2\perp} -
{\bm Q}_\perp)} e^{- \frac{a^2k_{3\perp}^2}{4}}.
\end{array}
\end{equation}
The Fourier transform of $\psi(z)^2$ will be called $\tilde\psi(k_z)$ for now, and evaluated explicitly later.
\begin{equation}
\arraycolsep 0.3 ex
\begin{array}{rl}
\displaystyle \psi(z_1)^2 = & \displaystyle \frac{1}{2\pi}\int_{-\infty}^\infty dk_{1z} \, e^{-ik_{1z}z} \tilde\psi(k_{1z}) \\ [3ex]
\displaystyle \psi(z_2)^2 = & \displaystyle \frac{1}{2\pi}\int_{-\infty}^\infty dk_{3z} \, e^{-ik_{3z}z} \tilde\psi(k_{3z}).
\end{array}
\end{equation}
The Fourier transform $\tilde\psi(k_z)$ is given by
\begin{equation}
\arraycolsep 0.3 ex
\begin{array}{rl}
\displaystyle \tilde\psi(k_z) = & \displaystyle N^2z_0^2\int_{-\infty}^0 dz \, e^{(k_b + ik_z)z} + N^2 \int_0^\infty dz \, (z + z_0)^2 \, e^{-(k_{Si} - ik_z)z} \\ [3ex]
= & \displaystyle \frac{N^2z_0^2}{(k_b + ik_z)} + \frac{N^2z_0^2}{(k_{Si} - ik_z)} + \frac{2N^2z_0} {(k_{Si} - ik_z)^2}+ \frac{2N^2}{(k_{Si} - ik_z)^3}.
\end{array}
\end{equation}
The integral $I ({\bm P}, {\bm Q})$ becomes
\begin{equation}
\arraycolsep 0.3 ex
\begin{array}{rl}
\displaystyle I_\psi (\bm R)= & \displaystyle \int \!\!\!\!\! \int\!\!\!\!\! \int\!\!\!\!\! \int\!\!\!\!\! \int \frac{d^3r_1 d^3r_2d^3 k_1 d^3k_2
d^3k_3}{128\pi^8} \, \frac{e^{-i({\bm k}_1 + {\bm k}_2)\cdot{\bm r}_1}e^{-i ({\bm k}_3 - {\bm k}_2)\cdot{\bm r}_2}}{k_2^2} \, e^{i{\bm k}_{1\perp}\cdot
{\bm P}} e^{i{\bm k}_{3\perp} \cdot {\bm Q}} e^{- \frac{a^2k_{1\perp}^2}{4}} e^{- \frac{a^2k_{3\perp}^2}{4}}\tilde\psi(k_{1z})\tilde\psi(k_{3z})
\\ [3ex]
= & \displaystyle \frac{1}{2\pi^2} \int \frac{d^3 k_1}{k_1^2} \, e^{i{\bm k}_{1\perp} \cdot ({\bm P} - {\bm Q})} e^{- \frac{a^2k_{1\perp}^2}{2}} \tilde\psi(k_{1z})\tilde\psi(-k_{1z}),
\end{array}
\end{equation}
where ${\bm R} = {\bm P} - {\bm Q}$. Since $\psi(z)$ is real, $\tilde\psi(k_{1z})\tilde\psi(-k_{1z}) = |\tilde\psi(k_{1z})|^2$. The Coulomb tunneling parameter in our setup is
\begin{equation}
\arraycolsep 0.3 ex
\begin{array}{rl}
\displaystyle s = & \displaystyle \frac{e^2}{\epsilon} \, l \, I_\psi (X_0 \hat{\bm x}) = \frac{e^2l}{2\pi^2\epsilon} \int \frac{d^3 k}{k^2} \, e^{ik_\perp X_0 \cos\theta} e^{- \frac{a^2k_\perp^2}{2}} |\tilde\psi(k_z)|^2 = \frac{e^2l}{\pi\epsilon} \int \frac{dk_\perp dk_z \, k_\perp}{k_\perp^2 + k_z^2} \, J_0(k_\perp X_0) \, e^{- \frac{a^2k_\perp^2}{2}} |\tilde\psi(k_z)|^2,
\end{array}
\end{equation}
recalling that $\tilde{s} \approx s$. This integral can be evaluated numerically for specific values of $k_b$, $k_{Si}$.

\section{$\Delta_E$ integrals}
\label{sec:Int}

The four integrals required for the evaluation of $\Delta_E$ are
\begin{equation}
\begin{array}{rl}
\displaystyle z_0^2\int_{-\infty}^0 dz\, z \, e^{(k_b + iq_z) z} = & \displaystyle z_0^2\bigg[z \, \frac{e^{(k_b + iq_z) z}}{(k_b + iq_z)}\bigg]^0_{-\infty} - z_0^2\int_{-\infty}^0 dz\, \frac{e^{(k_b + iq_z) z}}{(k_b + iq_z)} = - z_0^2\bigg[\frac{e^{(k_b + iq_z) z}}{(k_b + iq_z)^2}\bigg]_{-\infty}^0 = - \frac{z_0^2}{(k_b + iq_z)^2} \\ [3ex]

\displaystyle z_0^2 \int_0^{\infty} dz\, z e^{- (k_{Si} - iq_z) z} = & \displaystyle \frac{z_0^2}{(k_{Si} - iq_z)^2} \\ [3ex]

\displaystyle 2z_0\int_0^{\infty} dz\, z^2 e^{- (k_{Si} - iq_z) z} = & \displaystyle 2z_0 \bigg\{ 2\int_0^{\infty} dz\, z \frac{e^{- (k_{Si} - iq_z) z}}{(k_{Si} - iq_z)} \bigg\} = \frac{4z_0}{(k_{Si} - iq_z)^3} \\ [3ex]

\displaystyle \int_0^{\infty} dz\, z^3 e^{- (k_{Si} - iq_z) z} = & \displaystyle \frac{3}{(k_{Si} - iq_z)} \int_0^{\infty} dz\, z^2 e^{- (k_{Si} - iq_z) z} = \frac{6}{(k_{Si} - iq_z)^4}.
\end{array}
\end{equation}

\end{widetext}


\begin{thebibliography}{100}
\expandafter\ifx\csname natexlab\endcsname\relax\def\natexlab#1{#1}\fi
\expandafter\ifx\csname bibnamefont\endcsname\relax
  \def\bibnamefont#1{#1}\fi
\expandafter\ifx\csname bibfnamefont\endcsname\relax
  \def\bibfnamefont#1{#1}\fi
\expandafter\ifx\csname citenamefont\endcsname\relax
  \def\citenamefont#1{#1}\fi
\expandafter\ifx\csname url\endcsname\relax
  \def\url#1{\texttt{#1}}\fi
\expandafter\ifx\csname urlprefix\endcsname\relax\def\urlprefix{URL }\fi
\providecommand{\bibinfo}[2]{#2}
\providecommand{\eprint}[2][]{\url{#2}}

\bibitem[{\citenamefont{Nielsen and Chuang}(2000)}]{Nielsen_Chuang}
\bibinfo{author}{\bibfnamefont{M.~A.} \bibnamefont{Nielsen}} \bibnamefont{and}
  \bibinfo{author}{\bibfnamefont{I.~L.} \bibnamefont{Chuang}},
  \emph{\bibinfo{title}{Quantum Computation and Quantum Information}}
  (\bibinfo{publisher}{Cambridge University Press},
  \bibinfo{address}{Cambridge, UK}, \bibinfo{year}{2000}).

\bibitem[{\citenamefont{Liu et~al.}(2010)\citenamefont{Liu, Yao, and
  Sham}}]{LYS_Opt_QC_AP10}
\bibinfo{author}{\bibfnamefont{R.-B.} \bibnamefont{Liu}},
  \bibinfo{author}{\bibfnamefont{W.}~\bibnamefont{Yao}}, \bibnamefont{and}
  \bibinfo{author}{\bibfnamefont{L.~J.} \bibnamefont{Sham}},
  \bibinfo{journal}{Adv. Phys.} \textbf{\bibinfo{volume}{59}},
  \bibinfo{pages}{703} (\bibinfo{year}{2010}).

\bibitem[{\citenamefont{Kane}(1998)}]{Kane_Nature98}
\bibinfo{author}{\bibfnamefont{B.~E.} \bibnamefont{Kane}},
  \bibinfo{journal}{Nature (London)} \textbf{\bibinfo{volume}{393}},
  \bibinfo{pages}{133} (\bibinfo{year}{1998}).

\bibitem[{\citenamefont{Loss and DiVincenzo}(1998)}]{Loss_PRA98}
\bibinfo{author}{\bibfnamefont{D.}~\bibnamefont{Loss}} \bibnamefont{and}
  \bibinfo{author}{\bibfnamefont{D.~P.} \bibnamefont{DiVincenzo}},
  \bibinfo{journal}{Phys. Rev. A} \textbf{\bibinfo{volume}{57}},
  \bibinfo{pages}{120} (\bibinfo{year}{1998}).

\bibitem[{\citenamefont{Burkard et~al.}(1999)\citenamefont{Burkard, Loss, and
  DiVincenzo}}]{Burkard_PRB99}
\bibinfo{author}{\bibfnamefont{G.}~\bibnamefont{Burkard}},
  \bibinfo{author}{\bibfnamefont{D.}~\bibnamefont{Loss}}, \bibnamefont{and}
  \bibinfo{author}{\bibfnamefont{D.~P.} \bibnamefont{DiVincenzo}},
  \bibinfo{journal}{Phys. Rev. B} \textbf{\bibinfo{volume}{59}},
  \bibinfo{pages}{2070} (\bibinfo{year}{1999}).

\bibitem[{\citenamefont{Feher}(1959)}]{Feher_PR59}
\bibinfo{author}{\bibfnamefont{G.}~\bibnamefont{Feher}},
  \bibinfo{journal}{Phys. Rev.} \textbf{\bibinfo{volume}{114}},
  \bibinfo{pages}{1219} (\bibinfo{year}{1959}).

\bibitem[{\citenamefont{Tyryshkin et~al.}(2003)\citenamefont{Tyryshkin, Lyon,
  Astashkin, and Raitsimring}}]{Tyryshkin_PRB03}
\bibinfo{author}{\bibfnamefont{A.~M.} \bibnamefont{Tyryshkin}},
  \bibinfo{author}{\bibfnamefont{S.~A.} \bibnamefont{Lyon}},
  \bibinfo{author}{\bibfnamefont{A.~V.} \bibnamefont{Astashkin}},
  \bibnamefont{and} \bibinfo{author}{\bibfnamefont{A.~M.}
  \bibnamefont{Raitsimring}}, \bibinfo{journal}{Phys. Rev. B}
  \textbf{\bibinfo{volume}{68}}, \bibinfo{pages}{193207}
  (\bibinfo{year}{2003}).

\bibitem[{\citenamefont{Abe et~al.}(2004)\citenamefont{Abe, Itoh, Isoya, and
  Yamasaki}}]{Abe_PRB04}
\bibinfo{author}{\bibfnamefont{E.}~\bibnamefont{Abe}},
  \bibinfo{author}{\bibfnamefont{K.~M.} \bibnamefont{Itoh}},
  \bibinfo{author}{\bibfnamefont{J.}~\bibnamefont{Isoya}}, \bibnamefont{and}
  \bibinfo{author}{\bibfnamefont{S.}~\bibnamefont{Yamasaki}},
  \bibinfo{journal}{Phys.\ Rev.\ B} \textbf{\bibinfo{volume}{70}},
  \bibinfo{pages}{033204} (\bibinfo{year}{2004}).

\bibitem[{\citenamefont{Tyryshkin et~al.}(2006)\citenamefont{Tyryshkin, Morton,
  Benjamin, Ardavan, Briggs, Ager, and Lyon}}]{Tyryshkin_JPC06}
\bibinfo{author}{\bibfnamefont{A.~M.} \bibnamefont{Tyryshkin}},
  \bibinfo{author}{\bibfnamefont{J.~J.~L.} \bibnamefont{Morton}},
  \bibinfo{author}{\bibfnamefont{S.~C.} \bibnamefont{Benjamin}},
  \bibinfo{author}{\bibfnamefont{A.}~\bibnamefont{Ardavan}},
  \bibinfo{author}{\bibfnamefont{G.~A.~D.} \bibnamefont{Briggs}},
  \bibinfo{author}{\bibfnamefont{J.~W.} \bibnamefont{Ager}}, \bibnamefont{and}
  \bibinfo{author}{\bibfnamefont{S.~A.} \bibnamefont{Lyon}},
  \bibinfo{journal}{J. Phys. Condens. Matter} \textbf{\bibinfo{volume}{18}},
  \bibinfo{pages}{S783} (\bibinfo{year}{2006}).

\bibitem[{\citenamefont{Hanson et~al.}(2007)\citenamefont{Hanson, Kouwenhoven,
  Petta, Tarucha, and Vandersypen}}]{Hanson_RMP07}
\bibinfo{author}{\bibfnamefont{R.}~\bibnamefont{Hanson}},
  \bibinfo{author}{\bibfnamefont{L.~P.} \bibnamefont{Kouwenhoven}},
  \bibinfo{author}{\bibfnamefont{J.~R.} \bibnamefont{Petta}},
  \bibinfo{author}{\bibfnamefont{S.}~\bibnamefont{Tarucha}}, \bibnamefont{and}
  \bibinfo{author}{\bibfnamefont{L.~M.~K.} \bibnamefont{Vandersypen}},
  \bibinfo{journal}{Rev.\ Mod.\ Phys.} \textbf{\bibinfo{volume}{79}},
  \bibinfo{pages}{1217} (\bibinfo{year}{2007}).

\bibitem[{\citenamefont{Bluhm et~al.}(to be published)\citenamefont{Bluhm,
  Foletti, Neder, Rudner, Mahalu, Umansky, and Yacoby}}]{Bluhm_LongCoh_10}
\bibinfo{author}{\bibfnamefont{H.}~\bibnamefont{Bluhm}},
  \bibinfo{author}{\bibfnamefont{S.}~\bibnamefont{Foletti}},
  \bibinfo{author}{\bibfnamefont{I.}~\bibnamefont{Neder}},
  \bibinfo{author}{\bibfnamefont{M.}~\bibnamefont{Rudner}},
  \bibinfo{author}{\bibfnamefont{D.}~\bibnamefont{Mahalu}},
  \bibinfo{author}{\bibfnamefont{V.}~\bibnamefont{Umansky}}, \bibnamefont{and}
  \bibinfo{author}{\bibfnamefont{A.}~\bibnamefont{Yacoby}},
  \bibinfo{journal}{arXiv:1005.2995}  (\bibinfo{year}{to be published}).

\bibitem[{\citenamefont{Tyryshkin et~al.}(to be
  published)\citenamefont{Tyryshkin, Tojo, Morton, Riemann, Abrosimov, Becker,
  Pohl, Schenkel, Thewalt, Itoh et~al.}}]{Tyryshkin_IsoPureSiDnr_T2Secs_11}
\bibinfo{author}{\bibfnamefont{A.~M.} \bibnamefont{Tyryshkin}},
  \bibinfo{author}{\bibfnamefont{S.}~\bibnamefont{Tojo}},
  \bibinfo{author}{\bibfnamefont{J.~J.~L.} \bibnamefont{Morton}},
  \bibinfo{author}{\bibfnamefont{H.}~\bibnamefont{Riemann}},
  \bibinfo{author}{\bibfnamefont{N.~V.} \bibnamefont{Abrosimov}},
  \bibinfo{author}{\bibfnamefont{P.}~\bibnamefont{Becker}},
  \bibinfo{author}{\bibfnamefont{H.-J.} \bibnamefont{Pohl}},
  \bibinfo{author}{\bibfnamefont{T.}~\bibnamefont{Schenkel}},
  \bibinfo{author}{\bibfnamefont{M.~L.~W.} \bibnamefont{Thewalt}},
  \bibinfo{author}{\bibfnamefont{K.~M.} \bibnamefont{Itoh}},
  \bibnamefont{et~al.}, \bibinfo{journal}{arXiv:1105.3772}  (\bibinfo{year}{to
  be published}).

\bibitem[{\citenamefont{Prada et~al.}(2008)\citenamefont{Prada, Blick, and
  Joynt}}]{Prada_PRB08}
\bibinfo{author}{\bibfnamefont{M.}~\bibnamefont{Prada}},
  \bibinfo{author}{\bibfnamefont{R.~H.} \bibnamefont{Blick}}, \bibnamefont{and}
  \bibinfo{author}{\bibfnamefont{R.}~\bibnamefont{Joynt}},
  \bibinfo{journal}{Phys.\ Rev.\ B} \textbf{\bibinfo{volume}{77}},
  \bibinfo{pages}{115438} (\bibinfo{year}{2008}).

\bibitem[{\citenamefont{Tahan and Joynt}(2005)}]{Tahan_PRB05}
\bibinfo{author}{\bibfnamefont{C.}~\bibnamefont{Tahan}} \bibnamefont{and}
  \bibinfo{author}{\bibfnamefont{R.}~\bibnamefont{Joynt}},
  \bibinfo{journal}{Phys.\ Rev.\ B} \textbf{\bibinfo{volume}{71}},
  \bibinfo{pages}{075315} (\bibinfo{year}{2005}).

\bibitem[{\citenamefont{Li and Dery}(2011)}]{Dery_Si_SO_10}
\bibinfo{author}{\bibfnamefont{P.}~\bibnamefont{Li}} \bibnamefont{and}
  \bibinfo{author}{\bibfnamefont{H.}~\bibnamefont{Dery}},
  \bibinfo{journal}{Phys.\ Rev.\ Lett.} \textbf{\bibinfo{volume}{107}},
  \bibinfo{pages}{107203} (\bibinfo{year}{2011}).

\bibitem[{\citenamefont{Witzel et~al.}(2007)\citenamefont{Witzel, Hu, and {Das
  Sarma}}}]{Witzel_AHF_PRB07}
\bibinfo{author}{\bibfnamefont{W.~M.} \bibnamefont{Witzel}},
  \bibinfo{author}{\bibfnamefont{X.}~\bibnamefont{Hu}}, \bibnamefont{and}
  \bibinfo{author}{\bibfnamefont{S.}~\bibnamefont{{Das Sarma}}},
  \bibinfo{journal}{Phys.\ Rev.\ B} \textbf{\bibinfo{volume}{76}},
  \bibinfo{pages}{035212} (\bibinfo{year}{2007}).

\bibitem[{\citenamefont{Lim et~al.}(2009{\natexlab{a}})\citenamefont{Lim,
  Zwanenburg, Huebl, Mottonen, Chan, Morello, and
  Dzurak}}]{Lim_SingleElectron_APL09}
\bibinfo{author}{\bibfnamefont{W.~H.} \bibnamefont{Lim}},
  \bibinfo{author}{\bibfnamefont{F.~A.} \bibnamefont{Zwanenburg}},
  \bibinfo{author}{\bibfnamefont{H.}~\bibnamefont{Huebl}},
  \bibinfo{author}{\bibfnamefont{M.}~\bibnamefont{Mottonen}},
  \bibinfo{author}{\bibfnamefont{K.~W.} \bibnamefont{Chan}},
  \bibinfo{author}{\bibfnamefont{A.}~\bibnamefont{Morello}}, \bibnamefont{and}
  \bibinfo{author}{\bibfnamefont{A.~S.} \bibnamefont{Dzurak}},
  \bibinfo{journal}{Appl.\ Phys.\ Lett.} \textbf{\bibinfo{volume}{95}},
  \bibinfo{pages}{242102} (\bibinfo{year}{2009}{\natexlab{a}}).

\bibitem[{\citenamefont{Nordberg et~al.}(2009)\citenamefont{Nordberg, Stalford,
  Young, {Ten Eyck}, Eng, Tracy, Childs, Wendt, Grubbs, Stevens
  et~al.}}]{Nordberg_APL09}
\bibinfo{author}{\bibfnamefont{E.~P.} \bibnamefont{Nordberg}},
  \bibinfo{author}{\bibfnamefont{H.~L.} \bibnamefont{Stalford}},
  \bibinfo{author}{\bibfnamefont{R.}~\bibnamefont{Young}},
  \bibinfo{author}{\bibfnamefont{G.~A.} \bibnamefont{{Ten Eyck}}},
  \bibinfo{author}{\bibfnamefont{K.}~\bibnamefont{Eng}},
  \bibinfo{author}{\bibfnamefont{L.~A.} \bibnamefont{Tracy}},
  \bibinfo{author}{\bibfnamefont{K.~D.} \bibnamefont{Childs}},
  \bibinfo{author}{\bibfnamefont{J.~R.} \bibnamefont{Wendt}},
  \bibinfo{author}{\bibfnamefont{R.~K.} \bibnamefont{Grubbs}},
  \bibinfo{author}{\bibfnamefont{J.}~\bibnamefont{Stevens}},
  \bibnamefont{et~al.}, \bibinfo{journal}{Appl.\ Phys.\ Lett.}
  \textbf{\bibinfo{volume}{95}}, \bibinfo{pages}{202102}
  (\bibinfo{year}{2009}).

\bibitem[{\citenamefont{Yuan et~al.}(2011)\citenamefont{Yuan, Pan, Yang,
  Gilheart, Chen, Savage, Lagally, Eriksson, and
  Rimberg}}]{Eriksson_QD_SC_APL11}
\bibinfo{author}{\bibfnamefont{M.}~\bibnamefont{Yuan}},
  \bibinfo{author}{\bibfnamefont{F.}~\bibnamefont{Pan}},
  \bibinfo{author}{\bibfnamefont{Z.}~\bibnamefont{Yang}},
  \bibinfo{author}{\bibfnamefont{T.~J.} \bibnamefont{Gilheart}},
  \bibinfo{author}{\bibfnamefont{F.}~\bibnamefont{Chen}},
  \bibinfo{author}{\bibfnamefont{D.~E.} \bibnamefont{Savage}},
  \bibinfo{author}{\bibfnamefont{M.~G.} \bibnamefont{Lagally}},
  \bibinfo{author}{\bibfnamefont{M.~A.} \bibnamefont{Eriksson}},
  \bibnamefont{and} \bibinfo{author}{\bibfnamefont{A.~J.}
  \bibnamefont{Rimberg}}, \bibinfo{journal}{Appl.\ Phys.\ Lett.}
  \textbf{\bibinfo{volume}{98}}, \bibinfo{pages}{142104}
  (\bibinfo{year}{2011}).

\bibitem[{\citenamefont{Lim et~al.}(2009{\natexlab{b}})\citenamefont{Lim,
  Huebl, van Beveren, Rubanov, Spizzirri, Angus, Clark, and
  Dzurak}}]{Lim_APL09}
\bibinfo{author}{\bibfnamefont{W.~H.} \bibnamefont{Lim}},
  \bibinfo{author}{\bibfnamefont{H.}~\bibnamefont{Huebl}},
  \bibinfo{author}{\bibfnamefont{L.~H.~W.} \bibnamefont{van Beveren}},
  \bibinfo{author}{\bibfnamefont{S.}~\bibnamefont{Rubanov}},
  \bibinfo{author}{\bibfnamefont{P.~G.} \bibnamefont{Spizzirri}},
  \bibinfo{author}{\bibfnamefont{S.~J.} \bibnamefont{Angus}},
  \bibinfo{author}{\bibfnamefont{R.~G.} \bibnamefont{Clark}}, \bibnamefont{and}
  \bibinfo{author}{\bibfnamefont{A.~S.} \bibnamefont{Dzurak}},
  \bibinfo{journal}{Appl.\ Phys.\ Lett.} \textbf{\bibinfo{volume}{94}},
  \bibinfo{pages}{173502} (\bibinfo{year}{2009}{\natexlab{b}}).

\bibitem[{\citenamefont{Hu et~al.}(2007)\citenamefont{Hu, Churchill, Reilly,
  Xiang, Lieber, and Marcus}}]{MarcusGroup_NatureNano07}
\bibinfo{author}{\bibfnamefont{Y.}~\bibnamefont{Hu}},
  \bibinfo{author}{\bibfnamefont{H.~O.~H.} \bibnamefont{Churchill}},
  \bibinfo{author}{\bibfnamefont{D.~J.} \bibnamefont{Reilly}},
  \bibinfo{author}{\bibfnamefont{J.}~\bibnamefont{Xiang}},
  \bibinfo{author}{\bibfnamefont{C.~M.} \bibnamefont{Lieber}},
  \bibnamefont{and} \bibinfo{author}{\bibfnamefont{C.~M.}
  \bibnamefont{Marcus}}, \bibinfo{journal}{Nat. Nano.}
  \textbf{\bibinfo{volume}{2}}, \bibinfo{pages}{622} (\bibinfo{year}{2007}).

\bibitem[{\citenamefont{Tracy et~al.}(2010)\citenamefont{Tracy, Nordberg,
  Young, Pinilla, Stalford, Eyck, Eng, Childs, Stevens, Lilly
  et~al.}}]{Tracy_MOSFET_TunableDQD_APL10}
\bibinfo{author}{\bibfnamefont{L.~A.} \bibnamefont{Tracy}},
  \bibinfo{author}{\bibfnamefont{E.~P.} \bibnamefont{Nordberg}},
  \bibinfo{author}{\bibfnamefont{R.~W.} \bibnamefont{Young}},
  \bibinfo{author}{\bibfnamefont{C.~B.} \bibnamefont{Pinilla}},
  \bibinfo{author}{\bibfnamefont{H.~L.} \bibnamefont{Stalford}},
  \bibinfo{author}{\bibfnamefont{G.~A.~T.} \bibnamefont{Eyck}},
  \bibinfo{author}{\bibfnamefont{K.}~\bibnamefont{Eng}},
  \bibinfo{author}{\bibfnamefont{K.~D.} \bibnamefont{Childs}},
  \bibinfo{author}{\bibfnamefont{J.}~\bibnamefont{Stevens}},
  \bibinfo{author}{\bibfnamefont{M.~P.} \bibnamefont{Lilly}},
  \bibnamefont{et~al.}, \bibinfo{journal}{Appl.\ Phys.\ Lett.}
  \textbf{\bibinfo{volume}{97}}, \bibinfo{pages}{192110}
  (\bibinfo{year}{2010}).

\bibitem[{\citenamefont{Liu et~al.}(2008)\citenamefont{Liu, Fujisawa, Ono,
  Inokawa, Fujiwara, Takashina, and Hirayama}}]{Liu_PRB08}
\bibinfo{author}{\bibfnamefont{H.~W.} \bibnamefont{Liu}},
  \bibinfo{author}{\bibfnamefont{T.}~\bibnamefont{Fujisawa}},
  \bibinfo{author}{\bibfnamefont{Y.}~\bibnamefont{Ono}},
  \bibinfo{author}{\bibfnamefont{H.}~\bibnamefont{Inokawa}},
  \bibinfo{author}{\bibfnamefont{A.}~\bibnamefont{Fujiwara}},
  \bibinfo{author}{\bibfnamefont{K.}~\bibnamefont{Takashina}},
  \bibnamefont{and} \bibinfo{author}{\bibfnamefont{Y.}~\bibnamefont{Hirayama}},
  \bibinfo{journal}{Phys.\ Rev.\ B} \textbf{\bibinfo{volume}{77}},
  \bibinfo{pages}{073310} (\bibinfo{year}{2008}).

\bibitem[{\citenamefont{Shaji et~al.}(2008)\citenamefont{Shaji, Simmons,
  Thalakulam, Klein, Qin, Luo, Savage, Lagally, Rimberg, Joynt
  et~al.}}]{Shaji_NP08}
\bibinfo{author}{\bibfnamefont{N.}~\bibnamefont{Shaji}},
  \bibinfo{author}{\bibfnamefont{C.~B.} \bibnamefont{Simmons}},
  \bibinfo{author}{\bibfnamefont{M.}~\bibnamefont{Thalakulam}},
  \bibinfo{author}{\bibfnamefont{L.~J.} \bibnamefont{Klein}},
  \bibinfo{author}{\bibfnamefont{H.}~\bibnamefont{Qin}},
  \bibinfo{author}{\bibfnamefont{H.}~\bibnamefont{Luo}},
  \bibinfo{author}{\bibfnamefont{D.~E.} \bibnamefont{Savage}},
  \bibinfo{author}{\bibfnamefont{M.~G.} \bibnamefont{Lagally}},
  \bibinfo{author}{\bibfnamefont{A.~J.} \bibnamefont{Rimberg}},
  \bibinfo{author}{\bibfnamefont{R.}~\bibnamefont{Joynt}},
  \bibnamefont{et~al.}, \bibinfo{journal}{Nat. Phys.}
  \textbf{\bibinfo{volume}{4}}, \bibinfo{pages}{540} (\bibinfo{year}{2008}).

\bibitem[{\citenamefont{Simmons et~al.}(to be published)\citenamefont{Simmons,
  Koh, Shaji, Thalakulam, Klein, Qin, Luo, Savage, Lagally, Rimberg
  et~al.}}]{Simmons_DQD_SpinBloc_LET_PRB10}
\bibinfo{author}{\bibfnamefont{C.~B.} \bibnamefont{Simmons}},
  \bibinfo{author}{\bibfnamefont{T.~S.} \bibnamefont{Koh}},
  \bibinfo{author}{\bibfnamefont{N.}~\bibnamefont{Shaji}},
  \bibinfo{author}{\bibfnamefont{M.}~\bibnamefont{Thalakulam}},
  \bibinfo{author}{\bibfnamefont{L.~J.} \bibnamefont{Klein}},
  \bibinfo{author}{\bibfnamefont{H.}~\bibnamefont{Qin}},
  \bibinfo{author}{\bibfnamefont{H.}~\bibnamefont{Luo}},
  \bibinfo{author}{\bibfnamefont{D.~E.} \bibnamefont{Savage}},
  \bibinfo{author}{\bibfnamefont{M.~G.} \bibnamefont{Lagally}},
  \bibinfo{author}{\bibfnamefont{A.~J.} \bibnamefont{Rimberg}},
  \bibnamefont{et~al.}, \bibinfo{journal}{arXiv:1008.5398}  (\bibinfo{year}{to
  be published}).

\bibitem[{\citenamefont{Lai et~al.}(2011)\citenamefont{Lai, Lim, Yang,
  Zwanenburg, Morello, and Dzurak}}]{Lai_SiDQD_SpinBlock_10}
\bibinfo{author}{\bibfnamefont{N.~S.} \bibnamefont{Lai}},
  \bibinfo{author}{\bibfnamefont{W.~H.} \bibnamefont{Lim}},
  \bibinfo{author}{\bibfnamefont{C.~H.} \bibnamefont{Yang}},
  \bibinfo{author}{\bibfnamefont{F.~A.} \bibnamefont{Zwanenburg}},
  \bibinfo{author}{\bibfnamefont{A.}~\bibnamefont{Morello}}, \bibnamefont{and}
  \bibinfo{author}{\bibfnamefont{A.~S.} \bibnamefont{Dzurak}},
  \bibinfo{journal}{Sci.~Rep.} \textbf{\bibinfo{volume}{doi:10.1038/srep00110}}
  (\bibinfo{year}{2011}).

\bibitem[{\citenamefont{Borselli
  et~al.}(2011{\natexlab{a}})\citenamefont{Borselli, Eng, Croke, Maune, Huang,
  Ross, Kiselev, Deelman, Alvarado-Rodriguez, Schmitz
  et~al.}}]{Borselli_SpinBloc_11}
\bibinfo{author}{\bibfnamefont{M.~G.} \bibnamefont{Borselli}},
  \bibinfo{author}{\bibfnamefont{K.}~\bibnamefont{Eng}},
  \bibinfo{author}{\bibfnamefont{E.~T.} \bibnamefont{Croke}},
  \bibinfo{author}{\bibfnamefont{B.~M.} \bibnamefont{Maune}},
  \bibinfo{author}{\bibfnamefont{B.}~\bibnamefont{Huang}},
  \bibinfo{author}{\bibfnamefont{R.~S.} \bibnamefont{Ross}},
  \bibinfo{author}{\bibfnamefont{A.~A.} \bibnamefont{Kiselev}},
  \bibinfo{author}{\bibfnamefont{P.~W.} \bibnamefont{Deelman}},
  \bibinfo{author}{\bibfnamefont{I.}~\bibnamefont{Alvarado-Rodriguez}},
  \bibinfo{author}{\bibfnamefont{A.~E.} \bibnamefont{Schmitz}},
  \bibnamefont{et~al.}, \bibinfo{journal}{Appl.\ Phys.\ Lett.}
  \textbf{\bibinfo{volume}{99}}, \bibinfo{pages}{063109}
  (\bibinfo{year}{2011}{\natexlab{a}}).

\bibitem[{\citenamefont{Hayes et~al.}(2009)\citenamefont{Hayes, Kiselev,
  Borselli, Bui, III, Deelman, Maune, Milosavljevic, Moon, Ross
  et~al.}}]{Hayes_09}
\bibinfo{author}{\bibfnamefont{R.~R.} \bibnamefont{Hayes}},
  \bibinfo{author}{\bibfnamefont{A.~A.} \bibnamefont{Kiselev}},
  \bibinfo{author}{\bibfnamefont{M.~G.} \bibnamefont{Borselli}},
  \bibinfo{author}{\bibfnamefont{S.~S.} \bibnamefont{Bui}},
  \bibinfo{author}{\bibfnamefont{E.~T.~C.} \bibnamefont{III}},
  \bibinfo{author}{\bibfnamefont{P.~W.} \bibnamefont{Deelman}},
  \bibinfo{author}{\bibfnamefont{B.~M.} \bibnamefont{Maune}},
  \bibinfo{author}{\bibfnamefont{I.}~\bibnamefont{Milosavljevic}},
  \bibinfo{author}{\bibfnamefont{J.-S.} \bibnamefont{Moon}},
  \bibinfo{author}{\bibfnamefont{R.~S.} \bibnamefont{Ross}},
  \bibnamefont{et~al.}, \bibinfo{journal}{arXiv:0908.0173}
  (\bibinfo{year}{2009}).

\bibitem[{\citenamefont{Xiao et~al.}(2010{\natexlab{a}})\citenamefont{Xiao,
  House, and Jiang}}]{Xiao_MOS_SpinRelax_PRL10}
\bibinfo{author}{\bibfnamefont{M.}~\bibnamefont{Xiao}},
  \bibinfo{author}{\bibfnamefont{M.~G.} \bibnamefont{House}}, \bibnamefont{and}
  \bibinfo{author}{\bibfnamefont{H.~W.} \bibnamefont{Jiang}},
  \bibinfo{journal}{Phys.\ Rev.\ Lett.} \textbf{\bibinfo{volume}{104}},
  \bibinfo{pages}{096801} (\bibinfo{year}{2010}{\natexlab{a}}).

\bibitem[{\citenamefont{Thalakulam et~al.}((to be
  published))\citenamefont{Thalakulam, Simmons, Bael, Rosemeyer, Savage,
  Lagally, Friesen, Coppersmith, and Eriksson}}]{Thalak_1glShot_10}
\bibinfo{author}{\bibfnamefont{M.}~\bibnamefont{Thalakulam}},
  \bibinfo{author}{\bibfnamefont{C.~B.} \bibnamefont{Simmons}},
  \bibinfo{author}{\bibfnamefont{B.~J.~V.} \bibnamefont{Bael}},
  \bibinfo{author}{\bibfnamefont{B.~M.} \bibnamefont{Rosemeyer}},
  \bibinfo{author}{\bibfnamefont{D.~E.} \bibnamefont{Savage}},
  \bibinfo{author}{\bibfnamefont{M.~G.} \bibnamefont{Lagally}},
  \bibinfo{author}{\bibfnamefont{M.}~\bibnamefont{Friesen}},
  \bibinfo{author}{\bibfnamefont{S.~N.} \bibnamefont{Coppersmith}},
  \bibnamefont{and} \bibinfo{author}{\bibfnamefont{M.~A.}
  \bibnamefont{Eriksson}}, \bibinfo{journal}{arXiv:1010.0972}
  (\bibinfo{year}{(to be published)}).

\bibitem[{\citenamefont{Morello et~al.}(2010)\citenamefont{Morello, Pla,
  Zwanenburg, Chan, Huebl, Mottonen, Nugroho, Yang, van Donkelaar, Alves
  et~al.}}]{Morello_1glShot_Nature10}
\bibinfo{author}{\bibfnamefont{A.}~\bibnamefont{Morello}},
  \bibinfo{author}{\bibfnamefont{J.~J.} \bibnamefont{Pla}},
  \bibinfo{author}{\bibfnamefont{F.~A.} \bibnamefont{Zwanenburg}},
  \bibinfo{author}{\bibfnamefont{K.~W.} \bibnamefont{Chan}},
  \bibinfo{author}{\bibfnamefont{H.}~\bibnamefont{Huebl}},
  \bibinfo{author}{\bibfnamefont{M.}~\bibnamefont{Mottonen}},
  \bibinfo{author}{\bibfnamefont{C.~D.} \bibnamefont{Nugroho}},
  \bibinfo{author}{\bibfnamefont{C.}~\bibnamefont{Yang}},
  \bibinfo{author}{\bibfnamefont{J.~A.} \bibnamefont{van Donkelaar}},
  \bibinfo{author}{\bibfnamefont{A.~D.~C.} \bibnamefont{Alves}},
  \bibnamefont{et~al.}, \bibinfo{journal}{Nature}
  \textbf{\bibinfo{volume}{467}}, \bibinfo{pages}{687} (\bibinfo{year}{2010}).

\bibitem[{\citenamefont{Simmons et~al.}(2011)\citenamefont{Simmons, Prance,
  Bael, Koh, Shi, Savage, Lagally, Joynt, Friesen, Coppersmith
  et~al.}}]{Eriksson_SiQbt_Loading_10}
\bibinfo{author}{\bibfnamefont{C.~B.} \bibnamefont{Simmons}},
  \bibinfo{author}{\bibfnamefont{J.~R.} \bibnamefont{Prance}},
  \bibinfo{author}{\bibfnamefont{B.~J.~V.} \bibnamefont{Bael}},
  \bibinfo{author}{\bibfnamefont{T.~S.} \bibnamefont{Koh}},
  \bibinfo{author}{\bibfnamefont{Z.}~\bibnamefont{Shi}},
  \bibinfo{author}{\bibfnamefont{D.~E.} \bibnamefont{Savage}},
  \bibinfo{author}{\bibfnamefont{M.~G.} \bibnamefont{Lagally}},
  \bibinfo{author}{\bibfnamefont{R.}~\bibnamefont{Joynt}},
  \bibinfo{author}{\bibfnamefont{M.}~\bibnamefont{Friesen}},
  \bibinfo{author}{\bibfnamefont{S.~N.} \bibnamefont{Coppersmith}},
  \bibnamefont{et~al.}, \bibinfo{journal}{Phys.\ Rev.\ Lett.}
  \textbf{\bibinfo{volume}{106}}, \bibinfo{pages}{156804}
  (\bibinfo{year}{2011}).

\bibitem[{\citenamefont{Thalakulam et~al.}(2010)\citenamefont{Thalakulam,
  Simmons, Rosemeyer, Savage, Lagally, Friesen, Coppersmith, and
  Eriksson}}]{Eriksson_APL10}
\bibinfo{author}{\bibfnamefont{M.}~\bibnamefont{Thalakulam}},
  \bibinfo{author}{\bibfnamefont{C.~B.} \bibnamefont{Simmons}},
  \bibinfo{author}{\bibfnamefont{B.~M.} \bibnamefont{Rosemeyer}},
  \bibinfo{author}{\bibfnamefont{D.~E.} \bibnamefont{Savage}},
  \bibinfo{author}{\bibfnamefont{M.~G.} \bibnamefont{Lagally}},
  \bibinfo{author}{\bibfnamefont{M.}~\bibnamefont{Friesen}},
  \bibinfo{author}{\bibfnamefont{S.~N.} \bibnamefont{Coppersmith}},
  \bibnamefont{and} \bibinfo{author}{\bibfnamefont{M.~A.}
  \bibnamefont{Eriksson}}, \bibinfo{journal}{Appl.\ Phys.\ Lett.}
  \textbf{\bibinfo{volume}{96}}, \bibinfo{pages}{183104}
  (\bibinfo{year}{2010}).

\bibitem[{\citenamefont{Chan et~al.}(2011)\citenamefont{Chan, Mottonen,
  Kemppinen, Lai, Tan, Lim, and Dzurak}}]{ChanDzurak_SiQD_Shuttle_APL11}
\bibinfo{author}{\bibfnamefont{K.~W.} \bibnamefont{Chan}},
  \bibinfo{author}{\bibfnamefont{M.}~\bibnamefont{Mottonen}},
  \bibinfo{author}{\bibfnamefont{A.}~\bibnamefont{Kemppinen}},
  \bibinfo{author}{\bibfnamefont{N.~S.} \bibnamefont{Lai}},
  \bibinfo{author}{\bibfnamefont{K.~Y.} \bibnamefont{Tan}},
  \bibinfo{author}{\bibfnamefont{W.~H.} \bibnamefont{Lim}}, \bibnamefont{and}
  \bibinfo{author}{\bibfnamefont{A.~S.} \bibnamefont{Dzurak}},
  \bibinfo{journal}{Appl.\ Phys.\ Lett.} \textbf{\bibinfo{volume}{98}},
  \bibinfo{pages}{212103} (\bibinfo{year}{2011}).

\bibitem[{\citenamefont{Wild et~al.}(to be published)\citenamefont{Wild,
  Sailer, N\"utzel, Abstreiter, Ludwig, and Bougeard}}]{Wild_Si/SiGeQD_10}
\bibinfo{author}{\bibfnamefont{A.}~\bibnamefont{Wild}},
  \bibinfo{author}{\bibfnamefont{J.}~\bibnamefont{Sailer}},
  \bibinfo{author}{\bibfnamefont{J.}~\bibnamefont{N\"utzel}},
  \bibinfo{author}{\bibfnamefont{G.}~\bibnamefont{Abstreiter}},
  \bibinfo{author}{\bibfnamefont{S.}~\bibnamefont{Ludwig}}, \bibnamefont{and}
  \bibinfo{author}{\bibfnamefont{D.}~\bibnamefont{Bougeard}},
  \bibinfo{journal}{arXiv:1007.2404}  (\bibinfo{year}{to be published}).

\bibitem[{\citenamefont{Vrijen et~al.}(2000)\citenamefont{Vrijen, Yablonovitch,
  Wang, Jiang, Balandin, Roychowdhury, Mor, and DiVincenzo}}]{Vrijen_PRA00}
\bibinfo{author}{\bibfnamefont{R.}~\bibnamefont{Vrijen}},
  \bibinfo{author}{\bibfnamefont{E.}~\bibnamefont{Yablonovitch}},
  \bibinfo{author}{\bibfnamefont{K.}~\bibnamefont{Wang}},
  \bibinfo{author}{\bibfnamefont{H.~W.} \bibnamefont{Jiang}},
  \bibinfo{author}{\bibfnamefont{A.}~\bibnamefont{Balandin}},
  \bibinfo{author}{\bibfnamefont{V.}~\bibnamefont{Roychowdhury}},
  \bibinfo{author}{\bibfnamefont{T.}~\bibnamefont{Mor}}, \bibnamefont{and}
  \bibinfo{author}{\bibfnamefont{D.}~\bibnamefont{DiVincenzo}},
  \bibinfo{journal}{Phys.\ Rev.\ A} \textbf{\bibinfo{volume}{62}},
  \bibinfo{pages}{012306} (\bibinfo{year}{2000}).

\bibitem[{\citenamefont{Andresen et~al.}(2007)\citenamefont{Andresen, Brenner,
  Wellard, Yang, Hopf, Escott, Clark, Dzurak, Jamieson,
  et~al.}}]{Andresen_NanoLett07}
\bibinfo{author}{\bibfnamefont{S.~E.~S.} \bibnamefont{Andresen}},
  \bibinfo{author}{\bibfnamefont{R.}~\bibnamefont{Brenner}},
  \bibinfo{author}{\bibfnamefont{C.~J.} \bibnamefont{Wellard}},
  \bibinfo{author}{\bibfnamefont{C.}~\bibnamefont{Yang}},
  \bibinfo{author}{\bibfnamefont{T.}~\bibnamefont{Hopf}},
  \bibinfo{author}{\bibfnamefont{C.~C.} \bibnamefont{Escott}},
  \bibinfo{author}{\bibfnamefont{R.~G.} \bibnamefont{Clark}},
  \bibinfo{author}{\bibfnamefont{A.~S.} \bibnamefont{Dzurak}},
  \bibinfo{author}{\bibfnamefont{D.~N.} \bibnamefont{Jamieson}}, ,
  \bibnamefont{et~al.}, \bibinfo{journal}{Nano Lett.}
  \textbf{\bibinfo{volume}{7}}, \bibinfo{pages}{2000} (\bibinfo{year}{2007}).

\bibitem[{\citenamefont{Mitic et~al.}(2008)\citenamefont{Mitic, Petersson,
  Cassidy, Starrett, Gauja, Ferguson, Yang, Jamieson, Clark, and
  Dzurak}}]{Mitic_Nanotech08}
\bibinfo{author}{\bibfnamefont{M.}~\bibnamefont{Mitic}},
  \bibinfo{author}{\bibfnamefont{K.~D.} \bibnamefont{Petersson}},
  \bibinfo{author}{\bibfnamefont{M.~C.} \bibnamefont{Cassidy}},
  \bibinfo{author}{\bibfnamefont{R.~P.} \bibnamefont{Starrett}},
  \bibinfo{author}{\bibfnamefont{E.}~\bibnamefont{Gauja}},
  \bibinfo{author}{\bibfnamefont{A.~J.} \bibnamefont{Ferguson}},
  \bibinfo{author}{\bibfnamefont{C.}~\bibnamefont{Yang}},
  \bibinfo{author}{\bibfnamefont{D.~N.} \bibnamefont{Jamieson}},
  \bibinfo{author}{\bibfnamefont{R.~G.} \bibnamefont{Clark}}, \bibnamefont{and}
  \bibinfo{author}{\bibfnamefont{A.~S.} \bibnamefont{Dzurak}},
  \bibinfo{journal}{Nanotechnology} \textbf{\bibinfo{volume}{19}},
  \bibinfo{pages}{265201} (\bibinfo{year}{2008}).

\bibitem[{\citenamefont{Kuljanishvili et~al.}(2008)\citenamefont{Kuljanishvili,
  Kayis, Harrison, Piermarocchi, Kaplanand, Tessmer, Pfeiffer, and
  West}}]{Kuljanishvili_NP08}
\bibinfo{author}{\bibfnamefont{I.}~\bibnamefont{Kuljanishvili}},
  \bibinfo{author}{\bibfnamefont{C.}~\bibnamefont{Kayis}},
  \bibinfo{author}{\bibfnamefont{J.~F.} \bibnamefont{Harrison}},
  \bibinfo{author}{\bibfnamefont{C.}~\bibnamefont{Piermarocchi}},
  \bibinfo{author}{\bibfnamefont{T.~A.} \bibnamefont{Kaplanand}},
  \bibinfo{author}{\bibfnamefont{S.~H.} \bibnamefont{Tessmer}},
  \bibinfo{author}{\bibfnamefont{L.~N.} \bibnamefont{Pfeiffer}},
  \bibnamefont{and} \bibinfo{author}{\bibfnamefont{K.~W.} \bibnamefont{West}},
  \bibinfo{journal}{Nat. Phys.} \textbf{\bibinfo{volume}{4}},
  \bibinfo{pages}{227} (\bibinfo{year}{2008}).

\bibitem[{\citenamefont{Lansbergen et~al.}(2008)\citenamefont{Lansbergen,
  Rahman, Wellard, Woo, Caro, Collaert, Biesemans, Klimeck, Hollenberg, and
  Rogge}}]{Lansbergen_NP08}
\bibinfo{author}{\bibfnamefont{G.~P.} \bibnamefont{Lansbergen}},
  \bibinfo{author}{\bibfnamefont{R.}~\bibnamefont{Rahman}},
  \bibinfo{author}{\bibfnamefont{C.~J.} \bibnamefont{Wellard}},
  \bibinfo{author}{\bibfnamefont{I.}~\bibnamefont{Woo}},
  \bibinfo{author}{\bibfnamefont{J.}~\bibnamefont{Caro}},
  \bibinfo{author}{\bibfnamefont{N.}~\bibnamefont{Collaert}},
  \bibinfo{author}{\bibfnamefont{S.}~\bibnamefont{Biesemans}},
  \bibinfo{author}{\bibfnamefont{G.}~\bibnamefont{Klimeck}},
  \bibinfo{author}{\bibfnamefont{L.~C.~L.} \bibnamefont{Hollenberg}},
  \bibnamefont{and} \bibinfo{author}{\bibfnamefont{S.}~\bibnamefont{Rogge}},
  \bibinfo{journal}{Nat. Phys.} \textbf{\bibinfo{volume}{4}},
  \bibinfo{pages}{656} (\bibinfo{year}{2008}).

\bibitem[{\citenamefont{Fuhrer et~al.}(2009)\citenamefont{Fuhrer, F{\"u}chsle,
  Reusch, Weber, and Simmons}}]{Fuhrer_NanoLett09}
\bibinfo{author}{\bibfnamefont{A.}~\bibnamefont{Fuhrer}},
  \bibinfo{author}{\bibfnamefont{M.}~\bibnamefont{F{\"u}chsle}},
  \bibinfo{author}{\bibfnamefont{T.~C.~G.} \bibnamefont{Reusch}},
  \bibinfo{author}{\bibfnamefont{B.}~\bibnamefont{Weber}}, \bibnamefont{and}
  \bibinfo{author}{\bibfnamefont{M.~Y.} \bibnamefont{Simmons}},
  \bibinfo{journal}{Nano Lett.} \textbf{\bibinfo{volume}{9}},
  \bibinfo{pages}{707} (\bibinfo{year}{2009}).

\bibitem[{\citenamefont{Stegner et~al.}(2006)\citenamefont{Stegner, Boehme,
  Huebl, Stutzmann, Lips, and Brandt}}]{Stegner_NP06}
\bibinfo{author}{\bibfnamefont{A.~R.} \bibnamefont{Stegner}},
  \bibinfo{author}{\bibfnamefont{C.}~\bibnamefont{Boehme}},
  \bibinfo{author}{\bibfnamefont{H.}~\bibnamefont{Huebl}},
  \bibinfo{author}{\bibfnamefont{M.}~\bibnamefont{Stutzmann}},
  \bibinfo{author}{\bibfnamefont{K.}~\bibnamefont{Lips}}, \bibnamefont{and}
  \bibinfo{author}{\bibfnamefont{M.~S.} \bibnamefont{Brandt}},
  \bibinfo{journal}{Nature Phys.} \textbf{\bibinfo{volume}{2}},
  \bibinfo{pages}{835} (\bibinfo{year}{2006}).

\bibitem[{\citenamefont{Calderon et~al.}(2010)\citenamefont{Calderon, Verduijn,
  Lansbergen, Tettamanzi, Rogge, and
  Koiller}}]{Calderon_Dnr_Crg_Mismatch_PRB10}
\bibinfo{author}{\bibfnamefont{M.}~\bibnamefont{Calderon}},
  \bibinfo{author}{\bibfnamefont{J.}~\bibnamefont{Verduijn}},
  \bibinfo{author}{\bibfnamefont{G.}~\bibnamefont{Lansbergen}},
  \bibinfo{author}{\bibfnamefont{G.}~\bibnamefont{Tettamanzi}},
  \bibinfo{author}{\bibfnamefont{S.}~\bibnamefont{Rogge}}, \bibnamefont{and}
  \bibinfo{author}{\bibfnamefont{B.}~\bibnamefont{Koiller}},
  \bibinfo{journal}{Phys.\ Rev.\ B} \textbf{\bibinfo{volume}{82}},
  \bibinfo{pages}{075317} (\bibinfo{year}{2010}).

\bibitem[{\citenamefont{Mottonen et~al.}(2010)\citenamefont{Mottonen, Tan,
  Chan, Zwanenburg, Lim, Escott, Pirkkalainen, Morello, Yang, van Donkelaar
  et~al.}}]{Mottonen_PRB10}
\bibinfo{author}{\bibfnamefont{M.}~\bibnamefont{Mottonen}},
  \bibinfo{author}{\bibfnamefont{K.~Y.} \bibnamefont{Tan}},
  \bibinfo{author}{\bibfnamefont{K.~W.} \bibnamefont{Chan}},
  \bibinfo{author}{\bibfnamefont{F.~A.} \bibnamefont{Zwanenburg}},
  \bibinfo{author}{\bibfnamefont{W.~H.} \bibnamefont{Lim}},
  \bibinfo{author}{\bibfnamefont{C.~C.} \bibnamefont{Escott}},
  \bibinfo{author}{\bibfnamefont{J.-M.} \bibnamefont{Pirkkalainen}},
  \bibinfo{author}{\bibfnamefont{A.}~\bibnamefont{Morello}},
  \bibinfo{author}{\bibfnamefont{C.}~\bibnamefont{Yang}},
  \bibinfo{author}{\bibfnamefont{J.~A.} \bibnamefont{van Donkelaar}},
  \bibnamefont{et~al.}, \bibinfo{journal}{Phys.\ Rev.\ B}
  \textbf{\bibinfo{volume}{81}}, \bibinfo{pages}{161304}
  (\bibinfo{year}{2010}).

\bibitem[{\citenamefont{{Willems van Beveren} et~al.}(to be
  published)\citenamefont{{Willems van Beveren}, Huebl, and
  Morello}}]{Morello_PSi_rfDMR_11}
\bibinfo{author}{\bibfnamefont{L.~H.} \bibnamefont{{Willems van Beveren}}},
  \bibinfo{author}{\bibfnamefont{H.}~\bibnamefont{Huebl}}, \bibnamefont{and}
  \bibinfo{author}{\bibfnamefont{A.}~\bibnamefont{Morello}},
  \bibinfo{journal}{arXiv:1105.1235}  (\bibinfo{year}{to be published}).

\bibitem[{\citenamefont{Fuechsle et~al.}(2010)\citenamefont{Fuechsle,
  Mahapatra, Zwanenburg, Friesen, Eriksson, and Simmons}}]{Fuechsle_NNano10}
\bibinfo{author}{\bibfnamefont{M.}~\bibnamefont{Fuechsle}},
  \bibinfo{author}{\bibfnamefont{S.}~\bibnamefont{Mahapatra}},
  \bibinfo{author}{\bibfnamefont{F.~A.} \bibnamefont{Zwanenburg}},
  \bibinfo{author}{\bibfnamefont{M.}~\bibnamefont{Friesen}},
  \bibinfo{author}{\bibfnamefont{M.~A.} \bibnamefont{Eriksson}},
  \bibnamefont{and} \bibinfo{author}{\bibfnamefont{M.~Y.}
  \bibnamefont{Simmons}}, \bibinfo{journal}{Nature Nano.}
  \textbf{\bibinfo{volume}{5}}, \bibinfo{pages}{502} (\bibinfo{year}{2010}).

\bibitem[{\citenamefont{Fuechsle et~al.}(2012)\citenamefont{Fuechsle, Miwa,
  Mahapatra, Ryu, Lee, Warschkow, Hollenberg, Klimeck, and
  Simmons}}]{Fuechsle_SAT_NNano12}
\bibinfo{author}{\bibfnamefont{M.}~\bibnamefont{Fuechsle}},
  \bibinfo{author}{\bibfnamefont{J.~A.} \bibnamefont{Miwa}},
  \bibinfo{author}{\bibfnamefont{S.}~\bibnamefont{Mahapatra}},
  \bibinfo{author}{\bibfnamefont{H.}~\bibnamefont{Ryu}},
  \bibinfo{author}{\bibfnamefont{S.}~\bibnamefont{Lee}},
  \bibinfo{author}{\bibfnamefont{O.}~\bibnamefont{Warschkow}},
  \bibinfo{author}{\bibfnamefont{L.~C.~L.} \bibnamefont{Hollenberg}},
  \bibinfo{author}{\bibfnamefont{G.}~\bibnamefont{Klimeck}}, \bibnamefont{and}
  \bibinfo{author}{\bibfnamefont{M.~Y.} \bibnamefont{Simmons}},
  \bibinfo{journal}{Nature Nano.} \textbf{\bibinfo{volume}{DOI:
  10.1038/NNANO.2012.21}} (\bibinfo{year}{2012}).

\bibitem[{\citenamefont{Lu et~al.}(to be published)\citenamefont{Lu, Hoehne,
  Stegner, Dreher, Huebl, Stutzmann, and Brandt}}]{LuBrandt_PSi_FID_11}
\bibinfo{author}{\bibfnamefont{J.}~\bibnamefont{Lu}},
  \bibinfo{author}{\bibfnamefont{F.}~\bibnamefont{Hoehne}},
  \bibinfo{author}{\bibfnamefont{A.~R.} \bibnamefont{Stegner}},
  \bibinfo{author}{\bibfnamefont{L.}~\bibnamefont{Dreher}},
  \bibinfo{author}{\bibfnamefont{H.}~\bibnamefont{Huebl}},
  \bibinfo{author}{\bibfnamefont{M.}~\bibnamefont{Stutzmann}},
  \bibnamefont{and} \bibinfo{author}{\bibfnamefont{M.~S.}
  \bibnamefont{Brandt}}, \bibinfo{journal}{arXiv:1102.1550}  (\bibinfo{year}{to
  be published}).

\bibitem[{\citenamefont{Li et~al.}(2010)\citenamefont{Li, Cywi\'nski, Culcer,
  Hu, and {Das Sarma}}}]{Qiuzi_PRB10}
\bibinfo{author}{\bibfnamefont{Q.}~\bibnamefont{Li}},
  \bibinfo{author}{\bibfnamefont{L.}~\bibnamefont{Cywi\'nski}},
  \bibinfo{author}{\bibfnamefont{D.}~\bibnamefont{Culcer}},
  \bibinfo{author}{\bibfnamefont{X.}~\bibnamefont{Hu}}, \bibnamefont{and}
  \bibinfo{author}{\bibfnamefont{S.}~\bibnamefont{{Das Sarma}}},
  \bibinfo{journal}{Phys.\ Rev.\ B} \textbf{\bibinfo{volume}{81}},
  \bibinfo{pages}{085313} (\bibinfo{year}{2010}).

\bibitem[{\citenamefont{Nielsen et~al.}(2010)\citenamefont{Nielsen, Muller, and
  Carroll}}]{Nielsen_09}
\bibinfo{author}{\bibfnamefont{E.}~\bibnamefont{Nielsen}},
  \bibinfo{author}{\bibfnamefont{R.~P.} \bibnamefont{Muller}},
  \bibnamefont{and} \bibinfo{author}{\bibfnamefont{M.~S.}
  \bibnamefont{Carroll}}, \bibinfo{journal}{Phys.\ Rev.\ B}
  \textbf{\bibinfo{volume}{82}}, \bibinfo{pages}{075319}
  (\bibinfo{year}{2010}).

\bibitem[{\citenamefont{Nielsen and Muller}(to be
  published)}]{Nielsen_Config_10}
\bibinfo{author}{\bibfnamefont{E.}~\bibnamefont{Nielsen}} \bibnamefont{and}
  \bibinfo{author}{\bibfnamefont{R.~P.} \bibnamefont{Muller}},
  \bibinfo{journal}{arXiv:1006.2735}  (\bibinfo{year}{to be published}).

\bibitem[{\citenamefont{{Das Sarma} et~al.}(2011)\citenamefont{{Das Sarma},
  Wang, and Yang}}]{SDS_SiQD_Hubbard_PRB11}
\bibinfo{author}{\bibfnamefont{S.}~\bibnamefont{{Das Sarma}}},
  \bibinfo{author}{\bibfnamefont{X.}~\bibnamefont{Wang}}, \bibnamefont{and}
  \bibinfo{author}{\bibfnamefont{S.}~\bibnamefont{Yang}},
  \bibinfo{journal}{Phys.\ Rev.\ B} \textbf{\bibinfo{volume}{83}},
  \bibinfo{pages}{235314} (\bibinfo{year}{2011}).

\bibitem[{\citenamefont{Ramon}(to be published)}]{Ramon_2Qbt_11}
\bibinfo{author}{\bibfnamefont{G.}~\bibnamefont{Ramon}},
  \bibinfo{journal}{arXiv:1107:2968}  (\bibinfo{year}{to be published}).

\bibitem[{\citenamefont{Raith et~al.}(2011)\citenamefont{Raith, Stano, and
  Fabian}}]{Raith_SiQD_1e_SpinRelax_PRB11}
\bibinfo{author}{\bibfnamefont{M.}~\bibnamefont{Raith}},
  \bibinfo{author}{\bibfnamefont{P.}~\bibnamefont{Stano}}, \bibnamefont{and}
  \bibinfo{author}{\bibfnamefont{J.}~\bibnamefont{Fabian}},
  \bibinfo{journal}{Phys.\ Rev.\ B} \textbf{\bibinfo{volume}{83}},
  \bibinfo{pages}{195318} (\bibinfo{year}{2011}).

\bibitem[{\citenamefont{Wang et~al.}(2010{\natexlab{a}})\citenamefont{Wang,
  Shen, Sun, and Wu}}]{Wang_SiQD_ST_Relax_PRB10}
\bibinfo{author}{\bibfnamefont{L.}~\bibnamefont{Wang}},
  \bibinfo{author}{\bibfnamefont{K.}~\bibnamefont{Shen}},
  \bibinfo{author}{\bibfnamefont{B.~Y.} \bibnamefont{Sun}}, \bibnamefont{and}
  \bibinfo{author}{\bibfnamefont{M.~W.} \bibnamefont{Wu}},
  \bibinfo{journal}{Phys.\ Rev.\ B} \textbf{\bibinfo{volume}{81}},
  \bibinfo{pages}{235326} (\bibinfo{year}{2010}{\natexlab{a}}).

\bibitem[{\citenamefont{Wang and Wu}(2011)}]{WangWu_SiDQD_ST_Relax_11}
\bibinfo{author}{\bibfnamefont{L.}~\bibnamefont{Wang}} \bibnamefont{and}
  \bibinfo{author}{\bibfnamefont{M.~W.} \bibnamefont{Wu}},
  \bibinfo{journal}{J.\ Appl.\ Phys.} \textbf{\bibinfo{volume}{110}},
  \bibinfo{pages}{043716} (\bibinfo{year}{2011}).

\bibitem[{\citenamefont{Borhani and Hu}(2010)}]{Borhani_2spin_Relax_PRB10}
\bibinfo{author}{\bibfnamefont{M.}~\bibnamefont{Borhani}} \bibnamefont{and}
  \bibinfo{author}{\bibfnamefont{X.}~\bibnamefont{Hu}},
  \bibinfo{journal}{Phys.\ Rev.\ B} \textbf{\bibinfo{volume}{82}},
  \bibinfo{pages}{241302(R)} (\bibinfo{year}{2010}).

\bibitem[{\citenamefont{Hu}(2011)}]{Hu_2spin_e-ph_PRB11}
\bibinfo{author}{\bibfnamefont{X.}~\bibnamefont{Hu}}, \bibinfo{journal}{Phys.\
  Rev.\ B} \textbf{\bibinfo{volume}{83}}, \bibinfo{pages}{165322}
  (\bibinfo{year}{2011}).

\bibitem[{\citenamefont{Culcer et~al.}(2009{\natexlab{a}})\citenamefont{Culcer,
  Hu, and {Das Sarma}}}]{Culcer_APL09}
\bibinfo{author}{\bibfnamefont{D.}~\bibnamefont{Culcer}},
  \bibinfo{author}{\bibfnamefont{X.}~\bibnamefont{Hu}}, \bibnamefont{and}
  \bibinfo{author}{\bibfnamefont{S.}~\bibnamefont{{Das Sarma}}},
  \bibinfo{journal}{Appl.\ Phys.\ Lett.} \textbf{\bibinfo{volume}{95}},
  \bibinfo{pages}{073102} (\bibinfo{year}{2009}{\natexlab{a}}).

\bibitem[{\citenamefont{Witzel et~al.}(2010)\citenamefont{Witzel, Carroll,
  Morello, Cywinski, and {Das Sarma}}}]{Witzel_EnrichedSi_Decoh_PRL10}
\bibinfo{author}{\bibfnamefont{W.~M.} \bibnamefont{Witzel}},
  \bibinfo{author}{\bibfnamefont{M.~S.} \bibnamefont{Carroll}},
  \bibinfo{author}{\bibfnamefont{A.}~\bibnamefont{Morello}},
  \bibinfo{author}{\bibfnamefont{L.}~\bibnamefont{Cywinski}}, \bibnamefont{and}
  \bibinfo{author}{\bibfnamefont{S.}~\bibnamefont{{Das Sarma}}},
  \bibinfo{journal}{Phys.\ Rev.\ Lett.} \textbf{\bibinfo{volume}{105}},
  \bibinfo{pages}{187602} (\bibinfo{year}{2010}).

\bibitem[{\citenamefont{Assali et~al.}(2011)\citenamefont{Assali, Petrilli,
  Capaz, Koiller, Hu, and {Das Sarma}}}]{Assali_Hyperfine_PRB11}
\bibinfo{author}{\bibfnamefont{L.~V.~C.} \bibnamefont{Assali}},
  \bibinfo{author}{\bibfnamefont{H.~M.} \bibnamefont{Petrilli}},
  \bibinfo{author}{\bibfnamefont{R.~B.} \bibnamefont{Capaz}},
  \bibinfo{author}{\bibfnamefont{B.}~\bibnamefont{Koiller}},
  \bibinfo{author}{\bibfnamefont{X.}~\bibnamefont{Hu}}, \bibnamefont{and}
  \bibinfo{author}{\bibfnamefont{S.}~\bibnamefont{{Das Sarma}}},
  \bibinfo{journal}{Phys.\ Rev.\ B} \textbf{\bibinfo{volume}{83}},
  \bibinfo{pages}{165301} (\bibinfo{year}{2011}).

\bibitem[{\citenamefont{Churchill et~al.}(2009)\citenamefont{Churchill,
  Kuemmeth, Harlow, Bestwick, Rashba, Flensberg, Stwertka, Taychatanapat,
  Watson, and Marcus}}]{Churchill_PRL09}
\bibinfo{author}{\bibfnamefont{H.~O.~H.} \bibnamefont{Churchill}},
  \bibinfo{author}{\bibfnamefont{F.}~\bibnamefont{Kuemmeth}},
  \bibinfo{author}{\bibfnamefont{J.~W.} \bibnamefont{Harlow}},
  \bibinfo{author}{\bibfnamefont{A.~J.} \bibnamefont{Bestwick}},
  \bibinfo{author}{\bibfnamefont{E.~I.} \bibnamefont{Rashba}},
  \bibinfo{author}{\bibfnamefont{K.}~\bibnamefont{Flensberg}},
  \bibinfo{author}{\bibfnamefont{C.~H.} \bibnamefont{Stwertka}},
  \bibinfo{author}{\bibfnamefont{T.}~\bibnamefont{Taychatanapat}},
  \bibinfo{author}{\bibfnamefont{S.~K.} \bibnamefont{Watson}},
  \bibnamefont{and} \bibinfo{author}{\bibfnamefont{C.~M.}
  \bibnamefont{Marcus}}, \bibinfo{journal}{Phys.\ Rev.\ Lett.}
  \textbf{\bibinfo{volume}{102}}, \bibinfo{pages}{166802}
  (\bibinfo{year}{2009}).

\bibitem[{\citenamefont{P{\'a}lyi and Burkard}(2009)}]{Palyi_PRB09}
\bibinfo{author}{\bibfnamefont{A.}~\bibnamefont{P{\'a}lyi}} \bibnamefont{and}
  \bibinfo{author}{\bibfnamefont{G.}~\bibnamefont{Burkard}},
  \bibinfo{journal}{Phys.\ Rev.\ B} \textbf{\bibinfo{volume}{80}},
  \bibinfo{pages}{201404} (\bibinfo{year}{2009}).

\bibitem[{\citenamefont{P{\'a}lyi and Burkard}(2010)}]{Palyi_PRB10}
\bibinfo{author}{\bibfnamefont{A.}~\bibnamefont{P{\'a}lyi}} \bibnamefont{and}
  \bibinfo{author}{\bibfnamefont{G.}~\bibnamefont{Burkard}},
  \bibinfo{journal}{Phys.\ Rev.\ B} \textbf{\bibinfo{volume}{82}},
  \bibinfo{pages}{155424} (\bibinfo{year}{2010}).

\bibitem[{\citenamefont{P\'alyi and Burkard}(2011)}]{Palyi_PRL11}
\bibinfo{author}{\bibfnamefont{A.}~\bibnamefont{P\'alyi}} \bibnamefont{and}
  \bibinfo{author}{\bibfnamefont{G.}~\bibnamefont{Burkard}},
  \bibinfo{journal}{Phys.\ Rev.\ Lett.} \textbf{\bibinfo{volume}{106}},
  \bibinfo{pages}{086801} (\bibinfo{year}{2011}).

\bibitem[{\citenamefont{Wang et~al.}(2010{\natexlab{b}})\citenamefont{Wang,
  Cao, Tu, Li, Zhou, Hao, Su, Guo, Guo, and Jiang}}]{Wang_GfnQDwSET_APL10}
\bibinfo{author}{\bibfnamefont{L.-J.} \bibnamefont{Wang}},
  \bibinfo{author}{\bibfnamefont{G.}~\bibnamefont{Cao}},
  \bibinfo{author}{\bibfnamefont{T.}~\bibnamefont{Tu}},
  \bibinfo{author}{\bibfnamefont{H.-O.} \bibnamefont{Li}},
  \bibinfo{author}{\bibfnamefont{C.}~\bibnamefont{Zhou}},
  \bibinfo{author}{\bibfnamefont{X.-J.} \bibnamefont{Hao}},
  \bibinfo{author}{\bibfnamefont{Z.}~\bibnamefont{Su}},
  \bibinfo{author}{\bibfnamefont{G.-C.} \bibnamefont{Guo}},
  \bibinfo{author}{\bibfnamefont{G.-P.} \bibnamefont{Guo}}, \bibnamefont{and}
  \bibinfo{author}{\bibfnamefont{H.-W.} \bibnamefont{Jiang}},
  \bibinfo{journal}{Appl.\ Phys.\ Lett.} \textbf{\bibinfo{volume}{97}},
  \bibinfo{pages}{262113} (\bibinfo{year}{2010}{\natexlab{b}}).

\bibitem[{\citenamefont{Flensberg and
  Marcus}(2010)}]{Flensberg_CNT_Bends_PRB10}
\bibinfo{author}{\bibfnamefont{K.}~\bibnamefont{Flensberg}} \bibnamefont{and}
  \bibinfo{author}{\bibfnamefont{C.~M.} \bibnamefont{Marcus}},
  \bibinfo{journal}{Phys.\ Rev.\ B} \textbf{\bibinfo{volume}{81}},
  \bibinfo{pages}{195418} (\bibinfo{year}{2010}).

\bibitem[{\citenamefont{Shi et~al.}(2010)\citenamefont{Shi, Rong, Xu, Wang, Wu,
  Chong, Peng, Kniepert, Schoenfeld, Harneit et~al.}}]{ShiDu_DJ_PRL10}
\bibinfo{author}{\bibfnamefont{F.}~\bibnamefont{Shi}},
  \bibinfo{author}{\bibfnamefont{X.}~\bibnamefont{Rong}},
  \bibinfo{author}{\bibfnamefont{N.}~\bibnamefont{Xu}},
  \bibinfo{author}{\bibfnamefont{Y.}~\bibnamefont{Wang}},
  \bibinfo{author}{\bibfnamefont{J.}~\bibnamefont{Wu}},
  \bibinfo{author}{\bibfnamefont{B.}~\bibnamefont{Chong}},
  \bibinfo{author}{\bibfnamefont{X.}~\bibnamefont{Peng}},
  \bibinfo{author}{\bibfnamefont{J.}~\bibnamefont{Kniepert}},
  \bibinfo{author}{\bibfnamefont{R.-S.} \bibnamefont{Schoenfeld}},
  \bibinfo{author}{\bibfnamefont{W.}~\bibnamefont{Harneit}},
  \bibnamefont{et~al.}, \bibinfo{journal}{Phys.\ Rev.\ Lett.}
  \textbf{\bibinfo{volume}{105}}, \bibinfo{pages}{040504}
  (\bibinfo{year}{2010}).

\bibitem[{\citenamefont{Mazzeo et~al.}(to be published)\citenamefont{Mazzeo,
  Yablonovitch, and Jiang}}]{Mazzeo_GeQD_10}
\bibinfo{author}{\bibfnamefont{G.}~\bibnamefont{Mazzeo}},
  \bibinfo{author}{\bibfnamefont{E.}~\bibnamefont{Yablonovitch}},
  \bibnamefont{and} \bibinfo{author}{\bibfnamefont{H.~W.} \bibnamefont{Jiang}},
  \bibinfo{journal}{arXiv:1008.5168}  (\bibinfo{year}{to be published}).

\bibitem[{\citenamefont{Ando et~al.}(1982)\citenamefont{Ando, Fowler, and
  Stern}}]{Ando_RMP82}
\bibinfo{author}{\bibfnamefont{T.}~\bibnamefont{Ando}},
  \bibinfo{author}{\bibfnamefont{A.~B.} \bibnamefont{Fowler}},
  \bibnamefont{and} \bibinfo{author}{\bibfnamefont{F.}~\bibnamefont{Stern}},
  \bibinfo{journal}{Rev. Mod. Phys.} \textbf{\bibinfo{volume}{54}},
  \bibinfo{pages}{437} (\bibinfo{year}{1982}).

\bibitem[{\citenamefont{Koiller et~al.}(2001)\citenamefont{Koiller, Hu, and
  {Das Sarma}}}]{Koiller_PRL01}
\bibinfo{author}{\bibfnamefont{B.}~\bibnamefont{Koiller}},
  \bibinfo{author}{\bibfnamefont{X.}~\bibnamefont{Hu}}, \bibnamefont{and}
  \bibinfo{author}{\bibfnamefont{S.}~\bibnamefont{{Das Sarma}}},
  \bibinfo{journal}{Phys. Rev. Lett.} \textbf{\bibinfo{volume}{88}},
  \bibinfo{pages}{027903} (\bibinfo{year}{2001}).

\bibitem[{\citenamefont{Wellard and Hollenberg}(2005)}]{Wellard_PRB05}
\bibinfo{author}{\bibfnamefont{C.~J.} \bibnamefont{Wellard}} \bibnamefont{and}
  \bibinfo{author}{\bibfnamefont{L.~C.~L.} \bibnamefont{Hollenberg}},
  \bibinfo{journal}{Phys.\ Rev.\ B} \textbf{\bibinfo{volume}{72}},
  \bibinfo{pages}{085202} (\bibinfo{year}{2005}).

\bibitem[{\citenamefont{Koiller et~al.}(2006)\citenamefont{Koiller, Hu, and
  {Das Sarma}}}]{Koiller_PRB06}
\bibinfo{author}{\bibfnamefont{B.}~\bibnamefont{Koiller}},
  \bibinfo{author}{\bibfnamefont{X.}~\bibnamefont{Hu}}, \bibnamefont{and}
  \bibinfo{author}{\bibfnamefont{S.}~\bibnamefont{{Das Sarma}}},
  \bibinfo{journal}{Phys.\ Rev.\ B} \textbf{\bibinfo{volume}{73}},
  \bibinfo{pages}{045319} (\bibinfo{year}{2006}).

\bibitem[{\citenamefont{Culcer et~al.}(2009{\natexlab{b}})\citenamefont{Culcer,
  Cywi{\'n}ski, Li, Hu, and {Das Sarma}}}]{Culcer_PRB09}
\bibinfo{author}{\bibfnamefont{D.}~\bibnamefont{Culcer}},
  \bibinfo{author}{\bibfnamefont{{\L}.}~\bibnamefont{Cywi{\'n}ski}},
  \bibinfo{author}{\bibfnamefont{Q.}~\bibnamefont{Li}},
  \bibinfo{author}{\bibfnamefont{X.}~\bibnamefont{Hu}}, \bibnamefont{and}
  \bibinfo{author}{\bibfnamefont{S.}~\bibnamefont{{Das Sarma}}},
  \bibinfo{journal}{Phys.\ Rev.\ B} \textbf{\bibinfo{volume}{80}},
  \bibinfo{pages}{205302} (\bibinfo{year}{2009}{\natexlab{b}}).

\bibitem[{\citenamefont{Culcer et~al.}(2010{\natexlab{a}})\citenamefont{Culcer,
  Cywi{\'n}ski, Li, Hu, and {Das Sarma}}}]{Culcer_PRB10}
\bibinfo{author}{\bibfnamefont{D.}~\bibnamefont{Culcer}},
  \bibinfo{author}{\bibfnamefont{L.}~\bibnamefont{Cywi{\'n}ski}},
  \bibinfo{author}{\bibfnamefont{Q.}~\bibnamefont{Li}},
  \bibinfo{author}{\bibfnamefont{X.}~\bibnamefont{Hu}}, \bibnamefont{and}
  \bibinfo{author}{\bibfnamefont{S.}~\bibnamefont{{Das Sarma}}},
  \bibinfo{journal}{Phys.\ Rev.\ B} \textbf{\bibinfo{volume}{82}},
  \bibinfo{pages}{155312} (\bibinfo{year}{2010}{\natexlab{a}}).

\bibitem[{\citenamefont{Lai et~al.}(2006)\citenamefont{Lai, Lu, Pan, Tsui,
  Lyon, Liu, Xie, M\"uhlberger, and Sch\"affler}}]{Lai_PRB06}
\bibinfo{author}{\bibfnamefont{K.}~\bibnamefont{Lai}},
  \bibinfo{author}{\bibfnamefont{T.~M.} \bibnamefont{Lu}},
  \bibinfo{author}{\bibfnamefont{W.}~\bibnamefont{Pan}},
  \bibinfo{author}{\bibfnamefont{D.~C.} \bibnamefont{Tsui}},
  \bibinfo{author}{\bibfnamefont{S.}~\bibnamefont{Lyon}},
  \bibinfo{author}{\bibfnamefont{J.}~\bibnamefont{Liu}},
  \bibinfo{author}{\bibfnamefont{Y.~H.} \bibnamefont{Xie}},
  \bibinfo{author}{\bibfnamefont{M.}~\bibnamefont{M\"uhlberger}},
  \bibnamefont{and}
  \bibinfo{author}{\bibfnamefont{F.}~\bibnamefont{Sch\"affler}},
  \bibinfo{journal}{Phys.\ Rev.\ B} \textbf{\bibinfo{volume}{73}},
  \bibinfo{pages}{161301} (\bibinfo{year}{2006}).

\bibitem[{\citenamefont{Takashina et~al.}(2006)\citenamefont{Takashina, Ono,
  Fujiwara, Takahashi, and Hirayama}}]{Takashina_PRL06}
\bibinfo{author}{\bibfnamefont{K.}~\bibnamefont{Takashina}},
  \bibinfo{author}{\bibfnamefont{Y.}~\bibnamefont{Ono}},
  \bibinfo{author}{\bibfnamefont{A.}~\bibnamefont{Fujiwara}},
  \bibinfo{author}{\bibfnamefont{Y.}~\bibnamefont{Takahashi}},
  \bibnamefont{and} \bibinfo{author}{\bibfnamefont{Y.}~\bibnamefont{Hirayama}},
  \bibinfo{journal}{Phys.\ Rev.\ Lett.} \textbf{\bibinfo{volume}{96}},
  \bibinfo{pages}{236801} (\bibinfo{year}{2006}).

\bibitem[{\citenamefont{Goswami et~al.}(2007)\citenamefont{Goswami, Slinker,
  Friesen, McGuire, Truitt, Tahan, Klein, Chu, Mooney, {van der Weide}
  et~al.}}]{Goswami_NP07}
\bibinfo{author}{\bibfnamefont{S.}~\bibnamefont{Goswami}},
  \bibinfo{author}{\bibfnamefont{K.~A.} \bibnamefont{Slinker}},
  \bibinfo{author}{\bibfnamefont{M.}~\bibnamefont{Friesen}},
  \bibinfo{author}{\bibfnamefont{L.~M.} \bibnamefont{McGuire}},
  \bibinfo{author}{\bibfnamefont{J.~L.} \bibnamefont{Truitt}},
  \bibinfo{author}{\bibfnamefont{C.}~\bibnamefont{Tahan}},
  \bibinfo{author}{\bibfnamefont{L.~J.} \bibnamefont{Klein}},
  \bibinfo{author}{\bibfnamefont{J.~O.} \bibnamefont{Chu}},
  \bibinfo{author}{\bibfnamefont{P.~M.} \bibnamefont{Mooney}},
  \bibinfo{author}{\bibfnamefont{D.~W.} \bibnamefont{{van der Weide}}},
  \bibnamefont{et~al.}, \bibinfo{journal}{Nat. Phys.}
  \textbf{\bibinfo{volume}{3}}, \bibinfo{pages}{41} (\bibinfo{year}{2007}).

\bibitem[{\citenamefont{Lim et~al.}(2011)\citenamefont{Lim, Yang, Zwanenburg,
  and Dzurak}}]{Lim_SiQD_SpinFill_NT11}
\bibinfo{author}{\bibfnamefont{W.~H.} \bibnamefont{Lim}},
  \bibinfo{author}{\bibfnamefont{C.~H.} \bibnamefont{Yang}},
  \bibinfo{author}{\bibfnamefont{F.~A.} \bibnamefont{Zwanenburg}},
  \bibnamefont{and} \bibinfo{author}{\bibfnamefont{A.~S.}
  \bibnamefont{Dzurak}}, \bibinfo{journal}{Nanotechnology}
  \textbf{\bibinfo{volume}{22}}, \bibinfo{pages}{335704}
  (\bibinfo{year}{2011}).

\bibitem[{\citenamefont{Xiao et~al.}(2010{\natexlab{b}})\citenamefont{Xiao,
  House, and Jiang}}]{Xiao_Valley_APL10}
\bibinfo{author}{\bibfnamefont{M.}~\bibnamefont{Xiao}},
  \bibinfo{author}{\bibfnamefont{M.~G.} \bibnamefont{House}}, \bibnamefont{and}
  \bibinfo{author}{\bibfnamefont{H.~W.} \bibnamefont{Jiang}},
  \bibinfo{journal}{Appl.\ Phys.\ Lett.} \textbf{\bibinfo{volume}{97}},
  \bibinfo{pages}{032103} (\bibinfo{year}{2010}{\natexlab{b}}).

\bibitem[{\citenamefont{Borselli
  et~al.}(2011{\natexlab{b}})\citenamefont{Borselli, Ross, Kiselev, Croke,
  Holabird, Deelman, Warren, Alvarado-Rodriguez, Milosavljevic, Ku
  et~al.}}]{Borselli_Valley_APL11}
\bibinfo{author}{\bibfnamefont{M.~G.} \bibnamefont{Borselli}},
  \bibinfo{author}{\bibfnamefont{R.~S.} \bibnamefont{Ross}},
  \bibinfo{author}{\bibfnamefont{A.~A.} \bibnamefont{Kiselev}},
  \bibinfo{author}{\bibfnamefont{E.~T.} \bibnamefont{Croke}},
  \bibinfo{author}{\bibfnamefont{K.~S.} \bibnamefont{Holabird}},
  \bibinfo{author}{\bibfnamefont{P.~W.} \bibnamefont{Deelman}},
  \bibinfo{author}{\bibfnamefont{L.~D.} \bibnamefont{Warren}},
  \bibinfo{author}{\bibfnamefont{I.}~\bibnamefont{Alvarado-Rodriguez}},
  \bibinfo{author}{\bibfnamefont{I.}~\bibnamefont{Milosavljevic}},
  \bibinfo{author}{\bibfnamefont{F.~C.} \bibnamefont{Ku}},
  \bibnamefont{et~al.}, \bibinfo{journal}{Appl.\ Phys.\ Lett.}
  \textbf{\bibinfo{volume}{98}}, \bibinfo{pages}{123118}
  (\bibinfo{year}{2011}{\natexlab{b}}).

\bibitem[{\citenamefont{Boykin et~al.}(2004{\natexlab{a}})\citenamefont{Boykin,
  Klimeck, Friesen, Coppersmith, von Allmen, Oyafuso, and Lee}}]{Friesen_PRB04}
\bibinfo{author}{\bibfnamefont{T.~B.} \bibnamefont{Boykin}},
  \bibinfo{author}{\bibfnamefont{G.}~\bibnamefont{Klimeck}},
  \bibinfo{author}{\bibfnamefont{M.}~\bibnamefont{Friesen}},
  \bibinfo{author}{\bibfnamefont{S.~N.} \bibnamefont{Coppersmith}},
  \bibinfo{author}{\bibfnamefont{P.}~\bibnamefont{von Allmen}},
  \bibinfo{author}{\bibfnamefont{F.}~\bibnamefont{Oyafuso}}, \bibnamefont{and}
  \bibinfo{author}{\bibfnamefont{S.}~\bibnamefont{Lee}},
  \bibinfo{journal}{Phys.\ Rev.\ B} \textbf{\bibinfo{volume}{70}},
  \bibinfo{pages}{165325} (\bibinfo{year}{2004}{\natexlab{a}}).

\bibitem[{\citenamefont{Friesen et~al.}(2007)\citenamefont{Friesen, Chutia,
  Tahan, and Coppersmith}}]{Friesen_PRB07}
\bibinfo{author}{\bibfnamefont{M.}~\bibnamefont{Friesen}},
  \bibinfo{author}{\bibfnamefont{S.}~\bibnamefont{Chutia}},
  \bibinfo{author}{\bibfnamefont{C.}~\bibnamefont{Tahan}}, \bibnamefont{and}
  \bibinfo{author}{\bibfnamefont{S.~N.} \bibnamefont{Coppersmith}},
  \bibinfo{journal}{Phys.\ Rev.\ B} \textbf{\bibinfo{volume}{75}},
  \bibinfo{pages}{115318} (\bibinfo{year}{2007}).

\bibitem[{\citenamefont{Friesen and Coppersmith}(2010)}]{Friesen_PRB10}
\bibinfo{author}{\bibfnamefont{M.}~\bibnamefont{Friesen}} \bibnamefont{and}
  \bibinfo{author}{\bibfnamefont{S.~N.} \bibnamefont{Coppersmith}},
  \bibinfo{journal}{Phys.\ Rev.\ B} \textbf{\bibinfo{volume}{81}},
  \bibinfo{pages}{115324} (\bibinfo{year}{2010}).

\bibitem[{\citenamefont{Boykin et~al.}(2004{\natexlab{b}})\citenamefont{Boykin,
  Klimeck, Eriksson, Friesen, Coppersmith, von Allmen, Oyafuso, and
  Lee}}]{Boykin_APL04}
\bibinfo{author}{\bibfnamefont{T.~B.} \bibnamefont{Boykin}},
  \bibinfo{author}{\bibfnamefont{G.}~\bibnamefont{Klimeck}},
  \bibinfo{author}{\bibfnamefont{M.~A.} \bibnamefont{Eriksson}},
  \bibinfo{author}{\bibfnamefont{M.}~\bibnamefont{Friesen}},
  \bibinfo{author}{\bibfnamefont{S.~N.} \bibnamefont{Coppersmith}},
  \bibinfo{author}{\bibfnamefont{P.}~\bibnamefont{von Allmen}},
  \bibinfo{author}{\bibfnamefont{F.}~\bibnamefont{Oyafuso}}, \bibnamefont{and}
  \bibinfo{author}{\bibfnamefont{S.}~\bibnamefont{Lee}},
  \bibinfo{journal}{Appl.~Phys.~Lett.} \textbf{\bibinfo{volume}{84}},
  \bibinfo{pages}{115} (\bibinfo{year}{2004}{\natexlab{b}}).

\bibitem[{\citenamefont{Saraiva et~al.}(2009)\citenamefont{Saraiva, Calderon,
  Hu, {Das Sarma}, and Koiller}}]{Saraiva_PRB09}
\bibinfo{author}{\bibfnamefont{A.}~\bibnamefont{Saraiva}},
  \bibinfo{author}{\bibfnamefont{M.}~\bibnamefont{Calderon}},
  \bibinfo{author}{\bibfnamefont{X.}~\bibnamefont{Hu}},
  \bibinfo{author}{\bibfnamefont{S.}~\bibnamefont{{Das Sarma}}},
  \bibnamefont{and} \bibinfo{author}{\bibfnamefont{B.}~\bibnamefont{Koiller}},
  \bibinfo{journal}{Phys.\ Rev.\ B} \textbf{\bibinfo{volume}{80}},
  \bibinfo{pages}{081305} (\bibinfo{year}{2009}).

\bibitem[{\citenamefont{Saraiva et~al.}(2010)\citenamefont{Saraiva, Koiller,
  and Friesen}}]{Saraiva_Extended_PRB10}
\bibinfo{author}{\bibfnamefont{A.~L.} \bibnamefont{Saraiva}},
  \bibinfo{author}{\bibfnamefont{B.}~\bibnamefont{Koiller}}, \bibnamefont{and}
  \bibinfo{author}{\bibfnamefont{M.}~\bibnamefont{Friesen}},
  \bibinfo{journal}{Phys.\ Rev.\ B} \textbf{\bibinfo{volume}{82}},
  \bibinfo{pages}{245314} (\bibinfo{year}{2010}).

\bibitem[{\citenamefont{Nestoklon et~al.}(2006)\citenamefont{Nestoklon, Golub,
  and Ivchenko}}]{Nestoklon_PRB06}
\bibinfo{author}{\bibfnamefont{M.~O.} \bibnamefont{Nestoklon}},
  \bibinfo{author}{\bibfnamefont{L.~E.} \bibnamefont{Golub}}, \bibnamefont{and}
  \bibinfo{author}{\bibfnamefont{E.~L.} \bibnamefont{Ivchenko}},
  \bibinfo{journal}{Phys.\ Rev.\ B} \textbf{\bibinfo{volume}{73}},
  \bibinfo{pages}{235334} (\bibinfo{year}{2006}).

\bibitem[{\citenamefont{Srinivasan et~al.}(2008)\citenamefont{Srinivasan,
  Klimeck, and Rokhinson}}]{Srinivasan_APL08}
\bibinfo{author}{\bibfnamefont{S.}~\bibnamefont{Srinivasan}},
  \bibinfo{author}{\bibfnamefont{G.}~\bibnamefont{Klimeck}}, \bibnamefont{and}
  \bibinfo{author}{\bibfnamefont{L.~P.} \bibnamefont{Rokhinson}},
  \bibinfo{journal}{Appl.\ Phys.\ Lett.} \textbf{\bibinfo{volume}{93}},
  \bibinfo{pages}{112102} (\bibinfo{year}{2008}).

\bibitem[{\citenamefont{Rahman et~al.}(to be published)\citenamefont{Rahman,
  Verduijn, Kharche, Lansbergen, Klimeck, Hollenberg, and
  Rogge}}]{Rahman_Si_EngineerVOS_11}
\bibinfo{author}{\bibfnamefont{R.}~\bibnamefont{Rahman}},
  \bibinfo{author}{\bibfnamefont{J.}~\bibnamefont{Verduijn}},
  \bibinfo{author}{\bibfnamefont{N.}~\bibnamefont{Kharche}},
  \bibinfo{author}{\bibfnamefont{G.~P.} \bibnamefont{Lansbergen}},
  \bibinfo{author}{\bibfnamefont{G.}~\bibnamefont{Klimeck}},
  \bibinfo{author}{\bibfnamefont{L.~C.~L.} \bibnamefont{Hollenberg}},
  \bibnamefont{and} \bibinfo{author}{\bibfnamefont{S.}~\bibnamefont{Rogge}},
  \bibinfo{journal}{arXiv:1102.5311}  (\bibinfo{year}{to be published}).

\bibitem[{\citenamefont{Culcer et~al.}(2010{\natexlab{b}})\citenamefont{Culcer,
  Hu, and {Das Sarma}}}]{Culcer_Roughness_PRB10}
\bibinfo{author}{\bibfnamefont{D.}~\bibnamefont{Culcer}},
  \bibinfo{author}{\bibfnamefont{X.}~\bibnamefont{Hu}}, \bibnamefont{and}
  \bibinfo{author}{\bibfnamefont{S.}~\bibnamefont{{Das Sarma}}},
  \bibinfo{journal}{Phys.\ Rev.\ B} \textbf{\bibinfo{volume}{82}},
  \bibinfo{pages}{205315} (\bibinfo{year}{2010}{\natexlab{b}}).

\bibitem[{\citenamefont{Culcer et~al.}(2012)\citenamefont{Culcer, Saraiva, Hu,
  Koiller, and {Das Sarma}}}]{Culcer_ValleyQubit_PRL12}
\bibinfo{author}{\bibfnamefont{D.}~\bibnamefont{Culcer}},
  \bibinfo{author}{\bibfnamefont{A.}~\bibnamefont{Saraiva}},
  \bibinfo{author}{\bibfnamefont{X.}~\bibnamefont{Hu}},
  \bibinfo{author}{\bibfnamefont{B.}~\bibnamefont{Koiller}}, \bibnamefont{and}
  \bibinfo{author}{\bibfnamefont{S.}~\bibnamefont{{Das Sarma}}},
  \bibinfo{journal}{Phys.\ Rev.\ Lett.} \textbf{\bibinfo{volume}{108}},
  \bibinfo{pages}{126804} (\bibinfo{year}{2012}).

\bibitem[{\citenamefont{Bastard}(1988)}]{Bastard}
\bibinfo{author}{\bibfnamefont{G.}~\bibnamefont{Bastard}},
  \emph{\bibinfo{title}{Wave Mechanics Applied to Semiconductor
  Heterostructures}} (\bibinfo{publisher}{Halsted}, \bibinfo{address}{New
  York}, \bibinfo{year}{1988}).

\bibitem[{\citenamefont{Escott et~al.}(2010)\citenamefont{Escott, Zwanenburg,
  and Morello}}]{Escott_ResTnl_Nanotech10}
\bibinfo{author}{\bibfnamefont{C.~C.} \bibnamefont{Escott}},
  \bibinfo{author}{\bibfnamefont{F.~A.} \bibnamefont{Zwanenburg}},
  \bibnamefont{and} \bibinfo{author}{\bibfnamefont{A.}~\bibnamefont{Morello}},
  \bibinfo{journal}{Nanotechnology} \textbf{\bibinfo{volume}{21}},
  \bibinfo{pages}{274018} (\bibinfo{year}{2010}).

\bibitem[{\citenamefont{Ribeiro et~al.}(2010)\citenamefont{Ribeiro, Petta, and
  Burkard}}]{Ribeiro_ST+_10}
\bibinfo{author}{\bibfnamefont{H.}~\bibnamefont{Ribeiro}},
  \bibinfo{author}{\bibfnamefont{J.~R.} \bibnamefont{Petta}}, \bibnamefont{and}
  \bibinfo{author}{\bibfnamefont{G.}~\bibnamefont{Burkard}},
  \bibinfo{journal}{Phys.\ Rev.\ B} \textbf{\bibinfo{volume}{82}},
  \bibinfo{pages}{115445} (\bibinfo{year}{2010}).

\bibitem[{\citenamefont{Zimmerman et~al.}(2008)\citenamefont{Zimmerman, Huber,
  Simonds, Hourdakis, Fujiwara, Ono, Takahashi, Inokawa, Furlan, and
  Keller}}]{Zimmerman_LongTermCOD_JAP08}
\bibinfo{author}{\bibfnamefont{N.~M.} \bibnamefont{Zimmerman}},
  \bibinfo{author}{\bibfnamefont{W.~H.} \bibnamefont{Huber}},
  \bibinfo{author}{\bibfnamefont{B.}~\bibnamefont{Simonds}},
  \bibinfo{author}{\bibfnamefont{E.}~\bibnamefont{Hourdakis}},
  \bibinfo{author}{\bibfnamefont{A.}~\bibnamefont{Fujiwara}},
  \bibinfo{author}{\bibfnamefont{Y.}~\bibnamefont{Ono}},
  \bibinfo{author}{\bibfnamefont{Y.}~\bibnamefont{Takahashi}},
  \bibinfo{author}{\bibfnamefont{H.}~\bibnamefont{Inokawa}},
  \bibinfo{author}{\bibfnamefont{M.}~\bibnamefont{Furlan}}, \bibnamefont{and}
  \bibinfo{author}{\bibfnamefont{M.~W.} \bibnamefont{Keller}},
  \bibinfo{journal}{JAP} \textbf{\bibinfo{volume}{104}},
  \bibinfo{pages}{033710} (\bibinfo{year}{2008}).

\bibitem[{\citenamefont{Barthel et~al.}(2010)\citenamefont{Barthel, Medford,
  Marcus, Hanson, and Gossard}}]{Barthel_InterfaceDD_PRL10}
\bibinfo{author}{\bibfnamefont{C.}~\bibnamefont{Barthel}},
  \bibinfo{author}{\bibfnamefont{J.}~\bibnamefont{Medford}},
  \bibinfo{author}{\bibfnamefont{C.~M.} \bibnamefont{Marcus}},
  \bibinfo{author}{\bibfnamefont{M.~P.} \bibnamefont{Hanson}},
  \bibnamefont{and} \bibinfo{author}{\bibfnamefont{A.~C.}
  \bibnamefont{Gossard}}, \bibinfo{journal}{Phys.\ Rev.\ Lett.}
  \textbf{\bibinfo{volume}{105}}, \bibinfo{pages}{266808}
  (\bibinfo{year}{2010}).

\bibitem[{\citenamefont{Wang et~al.}(2012)\citenamefont{Wang, Bishop, Kestner,
  Barnes, Sun, and {Das Sarma}}}]{Wang_CompPuls_12}
\bibinfo{author}{\bibfnamefont{X.}~\bibnamefont{Wang}},
  \bibinfo{author}{\bibfnamefont{L.~S.} \bibnamefont{Bishop}},
  \bibinfo{author}{\bibfnamefont{J.~P.} \bibnamefont{Kestner}},
  \bibinfo{author}{\bibfnamefont{E.}~\bibnamefont{Barnes}},
  \bibinfo{author}{\bibfnamefont{K.}~\bibnamefont{Sun}}, \bibnamefont{and}
  \bibinfo{author}{\bibfnamefont{S.}~\bibnamefont{{Das Sarma}}},
  \bibinfo{journal}{arXiv:1202.5032}  (\bibinfo{year}{2012}).

\bibitem[{\citenamefont{Lu et~al.}(2011)\citenamefont{Lu, Bishop, Pluym, Means,
  Kotula, Cederberg, Tracy, Dominguez, Lilly, and
  Carroll}}]{Lu_Si/SiGe_Buried_11}
\bibinfo{author}{\bibfnamefont{T.~M.} \bibnamefont{Lu}},
  \bibinfo{author}{\bibfnamefont{N.~C.} \bibnamefont{Bishop}},
  \bibinfo{author}{\bibfnamefont{T.}~\bibnamefont{Pluym}},
  \bibinfo{author}{\bibfnamefont{J.}~\bibnamefont{Means}},
  \bibinfo{author}{\bibfnamefont{P.~G.} \bibnamefont{Kotula}},
  \bibinfo{author}{\bibfnamefont{J.}~\bibnamefont{Cederberg}},
  \bibinfo{author}{\bibfnamefont{L.~A.} \bibnamefont{Tracy}},
  \bibinfo{author}{\bibfnamefont{J.}~\bibnamefont{Dominguez}},
  \bibinfo{author}{\bibfnamefont{M.~P.} \bibnamefont{Lilly}}, \bibnamefont{and}
  \bibinfo{author}{\bibfnamefont{M.~S.} \bibnamefont{Carroll}},
  \bibinfo{journal}{Appl.\ Phys.\ Lett.} \textbf{\bibinfo{volume}{99}},
  \bibinfo{pages}{043101} (\bibinfo{year}{2011}).

\bibitem[{\citenamefont{Laird et~al.}(2010)\citenamefont{Laird, Taylor,
  DiVincenzo, Marcus, Hanson, and Gossard}}]{Laird_ExchQbt_PRB10}
\bibinfo{author}{\bibfnamefont{E.~A.} \bibnamefont{Laird}},
  \bibinfo{author}{\bibfnamefont{J.~M.} \bibnamefont{Taylor}},
  \bibinfo{author}{\bibfnamefont{D.~P.} \bibnamefont{DiVincenzo}},
  \bibinfo{author}{\bibfnamefont{C.~M.} \bibnamefont{Marcus}},
  \bibinfo{author}{\bibfnamefont{M.~P.} \bibnamefont{Hanson}},
  \bibnamefont{and} \bibinfo{author}{\bibfnamefont{A.~C.}
  \bibnamefont{Gossard}}, \bibinfo{journal}{Phys.\ Rev.\ B}
  \textbf{\bibinfo{volume}{82}}, \bibinfo{pages}{075403}
  (\bibinfo{year}{2010}).

\end{thebibliography}

\end{document}